\documentclass[aps,twocolumn,prx,superscriptaddress,natlib,showpacs]{revtex4-1}
\usepackage{graphicx}
\usepackage{dcolumn}
\usepackage{bm}
\usepackage{color}
\usepackage{amssymb}
\usepackage{amsmath}
\usepackage{amscd} 

\begin{document}

\title{Viscoelastic subdiffusion in a random Gaussian environment}
\author{Igor Goychuk}
\email{igoychuk@uni-potsdam.de}
\affiliation{Institute of Physics and Astronomy, University of Potsdam, 
Karl-Liebknecht-Str. 24/25, 14476 Potsdam-Golm, Germany}

\date{\today}

\begin{abstract}

Viscoelastic subdiffusion governed by a fractional Langevin equation is studied numerically in a random Gaussian environment modeled by stationary Gaussian potentials with decaying spatial correlations. This anomalous diffusion is archetypal for living cells, where cytoplasm is known to be viscoelastic and a spatial disorder  also naturally emerges. We obtain some first important insights into it within a model one-dimensional study. Two basic types of potential correlations are studied: short-range exponentially decaying and algebraically slow decaying with an infinite correlation length,
both for a moderate (several $k_BT$, in the units of thermal energy), and strong (5-10 $k_BT$) disorder. For a moderate disorder,  
 it is shown that on the ensemble level viscoelastic subdiffusion can easily overcome the medium's disorder. Asymptotically, it is not distinguishable from the disorder-free subdiffusion. 
However, a strong scatter in single-trajectory averages is nevertheless seen even for a moderate disorder. It features a weak ergodicity breaking, which occurs on a very long yet transient time scale. Furthermore, for a strong disorder,  a very long transient regime of logarithmic, Sinai-type diffusion emerges. It can last longer and be faster in the absolute terms for weakly decaying correlations as compare with the short-range correlations. 
Residence time distributions in a finite spatial domain are of a generalized log-normal type and are reminiscent also of a stretched exponential distribution. They can be easily confused for power-law distributions in view of the observed weak ergodicity breaking. This suggests a revision of some experimental data and their interpretation. 

\end{abstract}


\maketitle

\section{Introduction}

The research field of anomalous diffusion and transport \cite{Shlesinger74,Scher75,Bouchaud1990,AvrahamHavlin,Hughes, BouchaudAnnPhys90,Bouchaud92,Metzler01,FractDyn2011,GoychukACP12,MetzlerPCCP} currently flourishes getting ever more experimental support and manifestations in such diverse research areas as transport processes in living cells and polymeric solutions \cite{MasonPRL,AmblardPRL,Saxton97,Waigh,Seisenberg,Caspi,WeissBJ04,Tolic,BanksBJ05,Barkai,Bruno11,BrunoPRE,Golding,Guigas,SzymanskiPRL,HoflingReview,Jeon11,Luby,PanPRL,WongPRL,Harrison,Parry,Robert,Bruno11,BrunoPRE,PLoSONE14,PCCP14,PhysBio15,TabeiPNAS,WeigelPNAS,WeissPRE13,Santamaria,Bertzeva12}, colloidal systems \cite{MasonPRL,Waigh,EversReview,HanesPRE}, dust plasmas \cite{NukomuraPRL}, organic photoconductors \cite{SchubertPRB13}, conformational diffusion in proteins \cite{Yang03,KouPRL,GoychukPRE04,Kneller04,MinPRL,Calladrini10,Calligari11,Calligari15,GoychukPRE15,HuNatPhys16}, self-diffusion in lipid bilayers \cite{KnellerJCP11,JeonPRL12,JeonPRX}, diffusion of proteins on DNA strands \cite{Kong16,Kong17,Liu17},  to name just a few. 
Differently from normal diffusion, $\alpha=1$, the variance of the diffusing particle positions, $\langle \delta x^2(t)\rangle\propto t^\alpha$, often grows sublinearly, $\alpha<1$, or superlinearly, $\alpha>1$, in time, with some power law exponent $\alpha$. Accordingly, anomalous diffusion is classified into the subdiffusion, $\alpha<1$, and the superdiffusion, $\alpha>1$. This classification is, however, not complete. For example, Sinai diffusion \cite{Sinai82,Bouchaud1990,AvrahamHavlin} is characterized by a logarithmically slow growth, $\langle \delta x^2(t)\rangle\propto \ln^4 t $. This is clearly a sub-diffusion, which sometimes is named ultraslow \cite{Bouchaud1990,AvrahamHavlin}. 
The research field of anomalous diffusion remains rather controversial  because one and the same phenomena are often described by very different theories \cite{GoychukACP12,GoychukFractDyn11}, such as continuous time random walks (CTRWs)  with a divergent mean residence time in local traps \cite{Shlesinger74,Scher75,Bouchaud1990,AvrahamHavlin,Hughes,Metzler01}, and generalized Langevin dynamics with sub-Ohmic memory friction \cite{WeissBook,Pottier,Kupferman,GoychukPRE09,GoychukACP12}.
Such
different theories of fractional diffusion and transport can look at first very similar \cite{GoychukFractDyn11}.
However, a deeper analysis reveals fundamental differences in  ergodic \textit{vs.} weakly non-ergodic behavior \cite{Bouchaud92,MetzlerPCCP}, major features of nonlinear diffusion and transport in tilted periodic potentials \cite{GoychukPRE06Rapid,HeinsaluPRE06,GoychukPRE09,GoychukACP12,GoychukFractDyn11}, as well as in response to external time-periodic modulations \cite{BarbiPRL05,SokolovPRL06,HeinsaluPRL07,HeinsaluPRE09,GoychukPRE07Rapid,GoychukACP12,GoychukCTP14}. 

The pertinent diffusion in cytosol of biological cells is three-dimensional, and one in the cell plasma membrane is two-dimensional. However, the insights obtained from simplified one-dimensional theoretical models proved their usefulness over the years of research \cite{Shlesinger74,Scher75,Bouchaud1990,AvrahamHavlin,Hughes,Metzler01}. Hence, in this paper we will concentrate on a very simplified, minimal 1d model, which, nevertheless, is rich and complex enough. It suits well for getting such important insights and is based on 
two theoretical approaches to anomalous diffusion, which are especially important in view of their profound dynamical origin. 
One is based on the Bogolyubov-Ford-Kac-Mazur-Kubo-Zwanzig \cite{Bogolyubov,Ford65,Kubo66,Zwanzig73} generalized Langevin equation (GLE) 
\begin{equation}\label{GLEA}
m\ddot x+\int_{0}^{t}\eta(t-t')\dot x(t')dt'=f(x,t)+\xi(t),
\end{equation}
for the particle position $x(t)$ with an algebraically decaying memory kernel $\eta(t)\propto t^{-\alpha}$ \cite{WangTokuyama,LutzFLE,Pottier,GoychukPRL07,GoychukPRE09,GoychukACP12}.
Another one relies on normal, memory-less Langevin diffusion in random potentials \cite{Bouchaud1990,BouchaudAnnPhys90}. 
In Eq. (\ref{GLEA}), $m$ is the mass of the particle, $f(x,t)=-\partial U(x,t)/\partial x$ is an external force acting on the particle, which can be random, both in space and in time, and $\xi(t)$ is an equilibrium thermal noise with zero mean value. It has a Gaussian statistics and hence is named Gaussian. As any zero-mean stationary Gaussian process, it is completely characterized by its autocorrelation function (ACF), $\langle \xi(t)\xi(t')\rangle$. This one
is related to the memory kernel $\eta(t)$ by the classical fluctuation-dissipation relation (FDR), $\langle \xi(t)\xi(t')\rangle=k_BT\eta(|t-t'|)$ \cite{Kubo66,Zwanzig73,WeissBook}, named also the second 
fluctuation-dissipation theorem (FDT) by Kubo \cite{Kubo66}. Here, $T$ is temperature and $k_B$ is 
the Boltzmann constant. Standard Langevin equation presents a particular memory-less case with $\eta(t)=2\eta_0\delta(t)$, where $\eta_0$ is a viscous friction coefficient, yielding $\eta_0 \dot x$ for the friction term in (\ref{GLEA}). 

There are many studies of GLE dynamics, both potential-free and in some regular potentials, as well as of normal Langevin dynamics in random potentials. 
However, viscoelastic GLE subdiffusion in random potentials presents currently a practically unexplored topic despite its obvious relevance for diffusion processes in cytosol of living cells and other inhomogeneous viscoelastic media. Only in a parabolic weakly corrugated trapping potential such a diffusion  was partially addressed recently \cite{DuanEPJB}. This is the main purpose of this paper to do the first systematic study of viscoelastic GLE subdiffusion in stationary Gaussian potentials $U(x)$ with decaying correlations. We shall investigate two such models of general interest: (i) with exponentially decaying correlations (Ornstein-Uhlenbeck process in space), and (ii) algebraically decaying correlations possessing no effective correlation length.

Starting from classical works by Bogolyubov \cite{Bogolyubov}, Ford, Kac, Mazur \cite{Ford65}, and Zwanzig \cite{Zwanzig73}, the GLE (\ref{GLEA}) has repeatedly been derived \cite{LindenbergSeshardi,Ford88,Kupferman,WeissBook,GoychukACP12} from a fully dynamical system, where the environment is modeled by a large system of harmonic oscillators forming a thermal bath. The only non-dynamical element, which enters this theory, is that the initial positions and momenta of those oscillators are canonically distributed at a given fixed temperature. In this respect, this dynamical theory of Brownian motion presents a precursor and companion of molecular dynamics \cite{AlderJCP,AlderPRL}, in a very simplified, model fashion. It is also easy to generalize towards quantum-mechanical Brownian motion \cite{Magalinskii59,Caldeira83,Ford88,WeissBook} and to nonlinear models of coupling between the Brownian particle and its linear environment \cite{LindenbergSeshardi,WeissBook}. The influence of the environment in this approach is fully characterized by its spectral density \cite{Caldeira83,Ford88,WeissBook}, $J(\omega)$. It yields the memory kernel as \cite{WeissBook,GoychukACP12} 
$\eta(t)=(2/\pi)\int_0^{\infty}d\omega J(\omega)\cos(\omega t)/\omega$. Here, a very insightful basic model is
$J(\omega)=\eta_\alpha|\sin(\pi\alpha/2)|\omega^\alpha\exp(-\omega/\omega_c) $\cite{Caldeira83,Ford88,WeissBook}.
In accordance with it, the environment is customarily classified by the low-frequency behavior of $J(\omega)$
as Ohmic ($\alpha=1$), sub-Ohmic ($0<\alpha<1$) and super-Ohmic ($\alpha>1$) \cite{WeissBook}. Here, $\eta_\alpha$ is a fractional friction coefficient and $\omega_c$ is a frequency cutoff. It must be present in any condensed medium beyond the continuous medium approximation, which is, however, often used. 
This spectral model yields \cite{GoychukACP12,SiegleEPL11}
 \begin{eqnarray}\label{reg} 
\eta(t)&=&\eta_{\alpha}\frac{|\sin(\pi\alpha/2)|}{\pi/2}\Gamma(\alpha)
{\rm Re}(it+1/\omega_c)^{-\alpha}\\
&=&\eta_{\alpha}\frac{|\sin(\pi\alpha/2)|}{\pi/2}\frac{\Gamma(\alpha)\omega_c^\alpha}{(1+\omega_c^2t^2)^{\alpha/2}}
\cos[\alpha\arctan(\omega_c t)], \nonumber\;
\end{eqnarray}
where $\Gamma(z)$ is special gamma-function. Asymptotically, in the limit $t\to\infty$, and for the potential-free diffusion,
this model yields $\langle \delta x^2(t)\rangle\propto t^\alpha$,  for $0<\alpha<2$.
 It covers both sub- and super-diffusion. For $\alpha>2$, diffusion is ballistic \cite{WeissBook}, $\langle \delta x^2(t)\rangle\propto t^2$. In the singular limit, with unbounded energy spectrum, $\omega_c\to\infty$, and in neglecting quantum effects, the Ohmic model leads exactly \cite{WeissBook} to the standard classical Langevin  equation with viscous friction and thermal white Gaussian noise, which are related by the FDR. 

By the same token, the sub-Ohmic model in the singular limit $\omega_c\to\infty$ leads to a subdiffusive fractional Langevin equation (FLE) \cite{MainardiFLE,LutzFLE,Pottier,Kupferman,GoychukPRL07,GoychukACP12}. It is a GLE (\ref{GLEA}) with an algebraically decaying memory kernel, $\eta(t)=\eta_{\alpha} t^{-\alpha}/\Gamma(1-\alpha)$.  The corresponding frictional term with memory in Eq. (\ref{GLEA}) can be abbreviated as $\eta_\alpha d^\alpha x/dt^\alpha$ \cite{Caputo67,Mainardi97,Mathai17} using the notion of Caputo fractional derivative, $\frac{{\rm d}^\alpha x}{{\rm d}t^\alpha}:=\int_0^t  (t-t')^{-\alpha} \dot x(t')dt'/\Gamma(1-\alpha)$. 
Thermal noise entering this FLE  is a fractional Gaussian noise (fGn) \cite{Mandelbrot68}. 
This model of subdiffusion and its further generalizations emerge naturally in the context of anomalous diffusion in complex viscoelastic media such as complex liquids, including dense polymeric solutions, dust plasmas, colloids,  etc., with a prominent application to anomalous diffusion in cytosol of biological cells, as well as in intrinsic conformational dynamics of proteins.

Likewise, the super-Ohmic model with $1<\alpha<2$ leads to a superdiffusive GLE with a sign-changing and mostly negative memory kernel \cite{MainardiFLE,GoychukACP12,SiegleEPL11}. Its integral is always positive and tends to zero with the upper limit of integration  (vanishing integral friction). Its absolute value also decays  algebraically slow asymptotically. All these features can be understood from Eq. (\ref{reg}). For $1<\alpha<2$, the sign changes at $t_c=\tan(\pi/(2\alpha))/\omega_c$. The limit $\omega_c\to \infty$ is singular, $t_c\to 0+$, and for $t>0$, $\eta(t) = -\eta_\alpha t^{-\alpha}/|\Gamma(1-\alpha)|$. It must be handled with care because $\eta(t)$ is not a function but distribution in this limit. Then, the corresponding integral term in (\ref{GLEA}) can be short-handed \cite{MainardiFLE,GoychukACP12,SiegleEPL11} as $\eta_{\alpha}\sideset{_{0}}{_t}{\mathop{\hat D}^{\alpha-1}} \dot x(t)$ using the notion of fractional Riemann-Liouville derivative \cite{Mathai17}, $\sideset{_{0}}{_t}{\mathop{\hat D}^{\gamma}}v(t):=\frac{1}{\Gamma(1-\gamma)}
\frac{d}{dt}\int_{0}^t dt' \frac{v(t')}{(t-t')^\gamma}$, with $\gamma=\alpha-1$ and $v=\dot x$.

Such a related FLE, describing, however, an asymptotically normal diffusion, captures hydrodynamic memory effects by Boussinesq and Basset \cite{LandauHydro}. It takes the form of Eq. (\ref{GLEA}) with the frictional term abbreviated as $\eta_0\dot x+\eta_{\alpha}\sideset{_{0}}{_t}{\mathop{\hat D}^{\alpha-1}} \dot x(t)$ and two corresponding FDR-related noise terms \cite{MainardiFLE}. These memory effects lead to  a famous long tail in the velocity ACF of Brownian particles even in simple fluids \cite{AlderPRL,MainardiFLE,SiegleEPL11,GoychukACP12}.
Experimental manifestations of such  effects for Brownian particles were found quite recently \cite{HuangNatPhys,FranoschNature}. 

The GLE approach naturally provides a dynamical underpinning and justification of the mathematical model of fractional Brownian motion (fBm) by Kolmogorov \cite{Kolmogorov,KolmogorovTrans}, Mandelbrot and van Ness \cite{Mandelbrot68} within the FLE description upon neglecting the inertial effects.
This dynamical origin and consistency with thermodynamics and equilibrium statistical physics for undriven dynamics make this approach superior to many others in the field of anomalous diffusion \cite{GoychukACP12}. 
One of its special advantages is that it allows to study nonlinear anomalous dynamics in external multistable potentials. For example, such bistable subdiffusive dynamics was studied in Ref. \cite{GoychukPRE09} with a prominent result that the residence time distributions (RTDs) in the potential wells are of the stretched exponential type. In fact, no genuine rate description is possible.
This is due to the presence of slow fluctuations in the medium such that \textit{relatively fast} escape events  from one potential well to another can occur on the background of very sluggish, quasi-static fluctuations of the environment. Such fluctuations 
make the whole setup very different from one of the classical rate theory. The latter one assumes that the intrawell relaxation occurs  much faster than the escape events. This basic assumption is fundamentally broken for viscoelastic subdiffusive escape, where the slow modes of viscoelastic environment result into a time-modulation of the escape rate formed by the \textit{relatively} faster relaxation modes.
However, a characteristic time scale of transitions nevertheless exists. 
It is clearly seen in the distribution of  the logarithmically-transformed residence times, where the maximum of distribution
is well reproduced \cite{GoychukPRE09,GoychukACP12,GoychukPRE15} from a rate expression of the non-Markovian rate theory \cite{GroteHynes,HanggiRevModPhys}. It depends in the Arrhenius manner on the height of potential barrier.  

Next, the viscoelastic GLE subdiffusion in a washboard potential was shown to be insensitive asymptotically to the presence of potential \cite{GoychukPRE09}. It  approaches gradually a free-subdiffusion limit \cite{GoychukPRE09,GoychukACP12,GoychukFractDyn11} for any barrier height. This astounding feature is in a sharp contrast with both the intuition and  well-known results on normal diffusion\cite{HanggiRevModPhys} and fractional Fokker-Planck (FFP) dynamics \cite{GoychukPRE06Rapid,HeinsaluPRE06} in washboard potentials. It was also demonstrated for  diffusion and transport in other tilted periodic potentials including ratchets potentials with broken space inversion symmetry \cite{GoychukACP12,GoychukFractDyn11}.
Within a quantum-mechanical setting, it has been proven exactly, however, for strictly sinusoidal potentials only \cite{WeissBook,ChenLebowitz}. In fact, it is not caused by quantum-mechanical effects at all, as one might possibly believe, and it is not restricted by sinusoidal potentials only \cite{GoychukACP12}.  In this respect, it is also important to mention that the potential-free viscoelastic diffusion is ergodic: the ensemble and single-trajectory averages coincide  \cite{GoychukPRE09}, in a sharp contrast e.g. with continuous time random walk (CTRW) semi-Markovian subdiffusion \cite{HePRL08,LubelskiPRL08,MetzlerPCCP}. However, imposing a periodic potential makes it transiently non-ergodic \cite{GoychukPRE09}. These earlier results are very important to understand some of the key findings of this study. 

Another important approach to anomalous diffusion is based on normal diffusion in random potentials \cite{Bouchaud1990,BouchaudAnnPhys90,Hughes}. Such 
a description naturally emerge in inhomogeneous disordered materials, including  viscoelastic cytosol of living cells as well.  This is also a very rich and versatile approach. For example, the model of exponentially distributed energy disorder with root-mean-square (rms) amplitude of fluctuations $\sigma$ leads in a mean-field approximation to CTRW subdiffusion with a power law RTD, $\psi(\tau)\propto \tau^{-1-\alpha}, 0<\alpha=k_BT/\sigma<1$,  in local traps \cite{Hughes}. It is featured by divergent mean residence times (MRTs) \cite{Scher75,Shlesinger74} and is (weakly) nonergodic \cite{Bouchaud92,Bel05,LubelskiPRL08,HePRL08,Barkai,Sokolov09}: The ensemble and trajectory averages  are radically different.

However, in random potentials 
presenting stationary Gaussian processes in space such a diffusion is asymptotically normal for any decaying  correlations in space\cite{GoychukPRL14}, i. e. $\alpha(t)\to 1$, for $t\to\infty$. Here, a prominent result by de Gennes, B\"assler, and Zwanzig holds on the renormalized normal diffusion coefficient, $D_{\rm ren}=D_0\exp[-\sigma^2/(k_BT)^2]$, where $D_0$ is the potential-free diffusion coefficient \cite{DeGennes75,BasslerPRL87,BasslerReview,ZwanzigPNAS}.
The same renormalization is valid for FFP dynamics in such potentials, $\langle \delta x^2(t)\rangle \propto D_{\rm ren}t^\alpha$, where $D_0$ must be treated as fractional diffusion coefficient, see in Appendix A.
  The corresponding temperature dependence is often measured in disordered materials \cite{HecksherNatPhys} and the Gaussian model of energy disorder is physically well justified in many cases, e.g. for diffusion of electrons and holes in organic photoconductors \cite{BasslerReview,DunlapPRL96}, colloidal particles in random laser fields \cite{EversReview,BewerungePRA,HanesPRE} and regulatory proteins on DNA tracks \cite{GerlandPNAS,LassigReview,SlutskyPRE,BenichouPRL09,GoychukPRL14}.
However, for a sufficiently strong disorder $\sigma >2k_BT$  long subdiffusive transients occur on a mesoscale \cite{Romero98,KhouryPRL,SimonPRE13,GoychukPRL14} with a time-dependent sub-diffusion exponent
 $\alpha(t)\propto \log(t)$ \cite{Goychuk2017}. For $\sigma > (4 - 5) k_BT$, this mesoscale subdiffusion can reach even the macroscale \cite{GoychukPRL14}, and $\alpha(t)$ can be nearly constant for a very long time \cite{GoychukPRL14,Goychuk2017}. Remarkably, in this regime it exhibits the same temperature dependence, 
$\alpha\propto k_BT/\sigma$, as in the case of exponential disorder, despite a very different physical mechanism \cite{Goychuk2017}.
Such transient subdiffusion also exhibits a strong scatter in single-trajectory averages\cite{GoychukPRL14}
featuring a weak ergodicity breaking \cite{LubelskiPRL08,HePRL08,Barkai,MetzlerPCCP,Krapf18,Dean16}. 

Gaussian disorder characterized by a stationary random force $f(x)$ or a random drift coefficient \cite{Sinai82} is generally very different from the stationary potential disorder. It is also very important in applications\cite{Bouchaud1990}. Here, the simplest model is given \cite{Bouchaud1990} by the uncorrelated force disorder, $\langle f(x)f(x')\rangle \propto \delta(x-x')$, which leads to a
 logarithmically slow subdiffusion \cite{Sinai82,Bouchaud1990}, $\langle \delta x^2(t)\rangle \propto \log^a(t)$ with $a=4$ \cite{Bouchaud1990}. It is named Sinai diffusion. The corresponding
Gaussian potential $U(x)$ is a non-stationary random process.  It presents an unbounded Brownian motion (Wiener process) occurring in space, rather than time.   If the potential presents a fBm in space, a generalized Sinai diffusion with $a\neq 4$ emerges \cite{OshaninPRL}. Astoundingly, a generalized Sinai diffusion emerges also transiently in stationary correlated Gaussian potentials for a sufficiently strong disorder, 
$\sigma > 5 k_BT$.  This has been shown recently for four different models of disorder correlations in Ref. \cite{Goychuk2017}, where also the genuine mechanism of the discussed transient subdiffusion has been clarified using a scaling theory argumentation.

This paper is devoted to overdamped viscoelastic subdiffusion in random environments modeled by stationary random potentials with Gaussian statistics.
The rest of the paper is organized as follows. In Sec. II, we formulate  the model and the numerical approach.
In Sec. III, we present the main results and their discussion. Finally, in Sec. IV, the main conclusions will be drawn.\\

\section{Model and numerical approach}

We consider one-dimensional viscoelastic subdiffusion governed by the following overdamped subdiffusive FLE 
\cite{GoychukACP12,PRE13,GoychukPRE15} 
\begin{equation}
\eta_0\frac{{\rm d} x}{{\rm d}t}+\eta_\alpha\frac{{\rm d}^\alpha x}{{\rm d}t^\alpha}=f(x)+
 \xi_0(t)+\xi_{\alpha}(t),
\label{GLE1}
\end{equation}
where $0<\alpha<1$.  
The particles are moving in a random potential $U(x)$ yielding static random force $f(x)=-d U(x)/dx$. 
They are also subjected to thermal Gaussian forces $\xi_0(t)$ and $\xi_{\alpha}(t)$, as well as a memoryless Stokes friction with the friction coefficient $\eta_0$ and a friction with memory or frequency-dependent friction, which is characterized by the fractional friction coefficient $\eta_{\alpha}$, as detailed in the Introduction. 
Thermal noises and the corresponding  frictional parts
are related by the FDT relations $\langle\xi_0(t)\xi_0(t')\rangle=2k_B T\eta_0\delta(t-t')$, and
$\langle\xi_\alpha(t)\xi_\alpha(t')\rangle=k_B T\eta_\alpha | t-t'|^{-\alpha}/\Gamma(1-\alpha)$, correspondingly.  This ensures statistical equilibrium description in the absence of driving forces \cite{Kubo66}.
The both noises, $\xi_\alpha(t)$ and $\xi_0(t)$, present singular stochastic processes with infinite variance.
In the language of spectral bath functions, this description corresponds to $J(\omega)=\eta_0\omega+\eta_\alpha|\sin(\pi\alpha/2)|\omega^\alpha$, i.e. a mixture of Ohmic and sub-Ohmic thermal baths \cite{WeissBook},
in the singular limit $\omega_c\to\infty$.
In the case of cytosol or complex  polymeric liquids, the Stokes friction accounts for the water component of solution, whereas the friction with memory is caused by various dissolved polymers forming e.g. actin meshwork.
We neglect the inertial effects in anomalous dynamics,  which can also be easily included \cite{GoychukPRE09,GoychukACP12}, because we wish to arrive at a largest possible time scale accessible in numerical simulations. Moreover, very often such effects can indeed be neglected on physical grounds, see Appendix B for a justification. Hydrodynamic memory effects are also neglected, as usually. 

The solution of Eq. (\ref{GLE1}) for free subdiffusion, $f(x)=0$, yields \cite{PRE13}
\begin{eqnarray}\label{exact}
\langle\delta x^2(t)\rangle =2D_{0} t E_{1-\alpha,2}[-(t/\tau_0)^{1-\alpha}],
\end{eqnarray}
where $E_{a,b}(z):=\sum_0^{\infty}z^n/\Gamma(an+b)$ is generalized Mittag-Leffler function \cite{Mathai17},
$D_0=k_BT/\eta_{0}$ is a normal diffusion coefficient, and $\tau_0=(\eta_0/\eta_{\alpha})^{1/(1-\alpha)}$
is a transient time constant. Initially, for $t\ll \tau_0$, diffusion is normal, $\langle\delta x^2(t)\rangle \approx 2D_{0} t$, and asymptotically, $t\gg \tau_0$, it is anomalously slow,
$\langle\delta x^2(t)\rangle \sim 2D_{\alpha} t^\alpha/\Gamma(1+\alpha)$, with anomalous diffusion coefficient
$D_\alpha=k_BT/\eta_{\alpha}$. In the particular case of $\alpha=1/2$,  Eq. (\ref{exact}) yields
\begin{eqnarray}\label{exact2}
\langle\delta x^2(t)\rangle &= & 2D_{1/2} \Bigg \{  2\sqrt{\frac{t}{\pi}} \nonumber \\
 &+ &  \sqrt{\tau_0}
\left [ e^{t/ \tau_0}{\rm erfc}\left (\sqrt{\frac{t}{\tau_0}}\right ) -1\right ]
 \Bigg \},  
\end{eqnarray}
which is used to test numerical solutions below.

\begin{figure}[h]
\centering
\includegraphics[height=6cm]{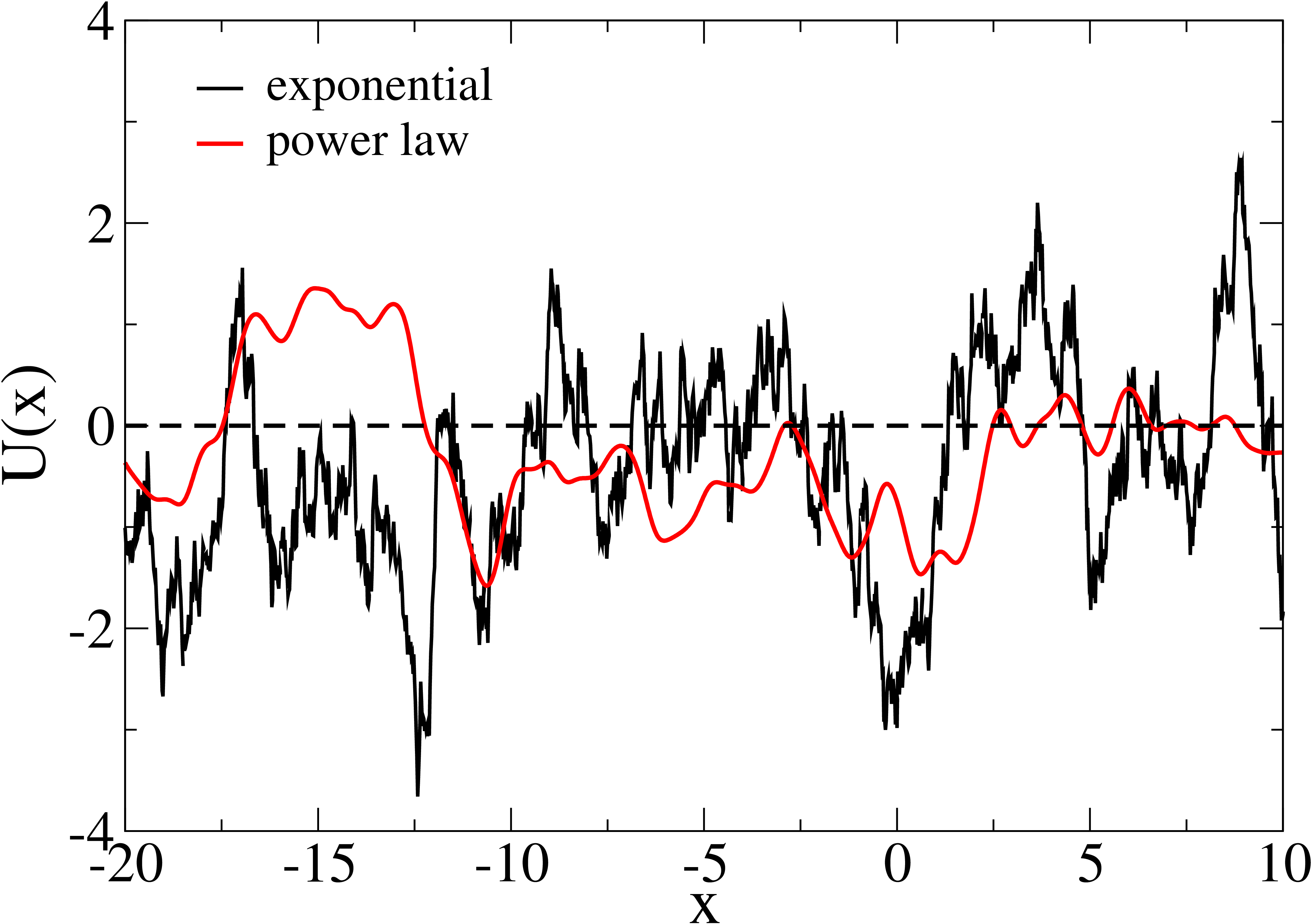}
\caption{(Color online) Realizations of random potentials (energy in units of $\sigma$, and coordinate in units of $\lambda$) for exponential and power law correlations. The lattice grid size is $\Delta x=0.02$.
In the case of power-law correlations, $\gamma=0.8$.}
\label{Fig1}       
\end{figure}

We consider stationary zero-mean random Gaussian potentials, which are completely characterized by the normalized stationary autocorrelation function, $g(x)=\langle U(x_0+x) U (x_0)\rangle /\langle U^2(x)\rangle$, and the rms of fluctuations, $\sigma=\langle U^2(x)\rangle^{1/2}$. Two models of correlations are considered: (i) exponential, $g(x)=\exp(-|x|/\lambda)$, and (ii) power law decaying, $g(x)=1/(1+x^2/\lambda^2)^{\gamma/2}$.  In the first case, $\lambda$ is the correlation length,
$\lambda_{\rm corr}=\int_0^\infty g(x)dx$, and in the second case 
$\lambda_{\rm corr}\to \infty$, for $0<\gamma <1$, as e.g. for diffusion of proteins on biological DNAs \cite{PengNature,AvrahamHavlin,Goychuk2017}. Then,
$\lambda$ is just a scaling parameter, which is convenient to use to scale distance in numerics. Furthermore, the time will be scaled in the units of $\tau_r=(\lambda^2\eta_\alpha/\sigma)^{1/\alpha}$,
$k_BT$ in units of $\sigma$, and normal friction coefficient $\eta_0$ in units of 
$\eta_{\alpha}\tau_r^{1-\alpha}$. 

\subsection{Numerical approach \\ \vspace{0cm}}

\subsubsection{Random potential generation}

Random Gaussian potentials are generated on a lattice evenly spaced with a grid size 
$\Delta x\ll \lambda$, using a spectral method in accordance with the numerical algorithm detailed in Ref. \cite{SimonFNL}. It requires to use a periodic boundary condition imposed on $U(x)$ with a very large period $L$. The method is based on the fact the power spectrum $S(k)$ of the random process $U(x)$ is obtained by the Fourier transform of its ACF (Wiener-Khinchine theorem). Moreover, it characterizes the absolute values of the amplitudes of the Fourier components $\hat U_k$ of $U(x)$ in the wave number space, $\langle \hat U_k\hat U_{k'}^*\rangle=L S(k)\delta_{k,k'}$ \cite{Papoulis}. First, $S(k)$ is obtained from $g(x)$ by a fast Fourier transform (FFT).
Next, Fourier transform of a Gaussian process is a Gaussian process \cite{Papoulis}. This allows to calculate the random wave amplitudes $\hat U_k$ from a set of independent Gaussian variables based on $S(k)$ and using another FFT. Finally, numerical inversion of the random wave components $\hat U_k$ to the coordinate space is done with inverse FFT. This yields the random realizations of $U(x)$. The method uses two direct and one inverse FFTs. The quality of the algorithm is checked and controlled by calculating numerically the ACF of the generated $U(x)$ and comparing it with the original ACF.
 It is impressively good. The readers are referred to Ref. \cite{SimonFNL} for further detail. 

In our numerics we fix $\Delta x=0.02$, and $L=2^{19}\approx 10^4$. Samples of random potentials with different correlations are presented in Fig. \ref{Fig1}. Notice a very rough character of the potential fluctuations for exponential correlations. There are many minima and maxima present within a correlation length. This is because this is a singular model of correlations. As a matter of fact, the corresponding force fluctuation $\langle \delta f^2(x)\rangle^{1/2}=\sqrt{2/(\Delta x \lambda)}\sigma$  diverges in the limit of 
$\Delta x\to 0$.   This is a crucial point: $\Delta x$ must be finite, on physical grounds in any such singular model \cite{Goychuk2017}. Otherwise, local \textit{static} forces acting on the particle can take very large values for a vanishing $\Delta x\to 0$. Any stochastic Langevin simulation in such a situation is damned to fail, if the time step $\Delta t$ in simulations is not chosen appropriately small: $\Delta t\to 0$ with $\Delta x\to 0$. The smaller $\Delta x$, the smaller  $\Delta t$ must be used for Langevin simulations of such singular models of disorder \cite{GoychukPRL14}.   The model of a delta-correlated potential is physically a model with $U(x)$ values uncorrelated on the lattice sites. However, because of continuity of potential it remains correlated between the sites of the lattice, anyway \cite{Goychuk2017}.
This is actually the case, where the potential fluctuations have the wildest character, and do not exhibit a \textit{local bias}, which otherwise is \textit{always present} because of correlations. In our numerics, we connect the lattice values of potential by parabolic splines, i.e. the potential is locally parabolic and 
$f(x)$ is piece-wise linear. Notice also that the power-law correlated potential is much smoother and it does not display the discussed singularity in the limit $\Delta x \to 0$. \\

\subsubsection{Approximation of the memory kernel}

Our numerical approach to integrate the FLE dynamics is based on approximation of the power-law memory kernel by a sum of exponentials,  
\begin{eqnarray}\label{Prony}
\eta(t)=\sum_{i=1}^N k_i\exp(-\nu_i t), 
\end{eqnarray}
i.e. using a Prony series expansion \cite{GoychukPRE09,McKinley}. Eq. (\ref{Prony}) presents a particular case of more general Prony series \cite{Prony,Hauer,Park99,Schapery99}, $s(t)=\sum_{i=1}^N k_i\exp(-\nu_i t)\cos(\omega_i t+\delta_i)$, used to approximate any empirical signal $s(t)$ using $N$ decaying wave-forms, with decay rates $\nu_i$, frequencies $\omega_i$, and phase shifts $\delta_i$. It presents a further generalization of Fourier series and has been introduced originally by Prony in 1795 \cite{Prony}. The expansions of viscoelastic memory kernels like one in Eq. (\ref{Prony}) naturally emerge in the theory of polymer dynamics and polymeric melts \cite{Doi}. For example, for the Rouse model of a polymer consisting of $N$ monomers \cite{Doi},
$\nu_i=i^p\nu_{l}$ with $p=2$ and $k_i=const$, in terms of some smallest $\nu_{ l}$ in the hierarchy of relaxation rates $\nu_i$. This yields \cite{McKinley} $\eta(t)\propto 1/t^{1/p}$, e.g.  $\alpha=1/2$ for $p=2$, in the range of $\tau_l\ll t \ll \tau_{h}$, with $\tau_{h}=1/\nu_{l}$, and $\tau_{l}=N^p/\nu_{l}=\nu_0$. Notice that it is $\nu_0$ which plays a fundamental role being related to the overdamped dynamics of one monomer in the Rouse chain \cite{Doi}. It determines the lower time cutoff of the power law dependence $\eta(t)\propto t^{-\alpha}$. Accordingly, $\nu_{l}=\nu_0/N^p$. The larger $N$, the larger is the upper time cutoff $\tau_h=N^p/\nu_0$, whereas $\tau_l$ remains unchanged. Notice that the both time cutoffs naturally emerge in the dynamics of polymeric melts. They always exist. 

The polymeric scaling of the relaxation rates $\nu_i$ is not unique. Another way is to choose a fractal scaling, $\nu_i=\nu_0/b^{i-1}$, with the spring constants 
$k_i =C_\alpha(b)\nu_i^\alpha/\Gamma(1-\alpha)\propto \nu_i^\alpha$, where $C_\alpha(b)$
is some constant, which depends on $\alpha$ and $b$ \cite{PalmerPRL85,Hughes,GoychukPRE09,GoychukACP12}.
 It is used e.g. in a phenomenological temporary network model of polymeric melts \cite{Larson}.
Already for a rather crude scaling with $b=10$, the accuracy of this approximation between two memory cutoffs, $\tau_l=1/\nu_0$ and $\tau_h=\tau_l b^{N-1}$, reaches several percents for $\alpha=0.5$ \cite{GoychukPRE09}. 
The great advantage of the fractal scaling over the polymeric one is that it is requires a much smaller number $N$ of viscoelastic modes in the memory kernel approximation. 
Indeed, for having the same range $\tau_h/\tau_l$ of power law scaling within the polymeric scaling as within the fractal scaling with $N$ terms one needs
\begin{eqnarray}
M=b^{(N-1)/p}\;
\end{eqnarray}
terms in (\ref{Prony}). For example, for $\alpha=0.5$ and $p=2$ this would give $M=10^5$ (!) instead of $N=11$ within the fractal scaling with $b=10$, or $N=35$ with $b=2$. This clearly establishes superiority of the fractal scaling in numerics \cite{GoychukACP12}.
It 
allows for a numerically very efficient approach to integrate FLE \cite{GoychukPRE09,GoychukACP12}. Notice that $\nu_0$ can be chosen somewhat smaller (to avoid numerical instability) than
the inverse time step $1/\Delta t$ in the numerical simulation, and even $N\sim 10 - 20$ is typically sufficient in numerical simulations with $b=10$. For the scaling with $b=2$, the accuracy
of the memory kernel approximation improves to $0.01\%$ \cite{MMNP13}. Then, however, one should also increase $N$ accordingly, which would provide an extra time burden in the numerics. Accuracy of several percents is normally sufficient. 

\subsubsection{Markovian embedding}

Next, we introduce  a subset of auxiliary  variables $x_i$ \cite{GoychukPRE09} corresponding to the viscoelastic modes of the environment. Physically, they can be interpreted as coordinates of some auxiliary Brownian quasi-particles modeling viscoelastic Maxwellian modes of the environment and elastically coupled with spring constants $k_i$ to the central Brownian particle \cite{GoychukACP12}. The fractional Gaussian noise 
$\eta_{\alpha}(t)$ in this approach is approximated by a sum of Ornstein-Uhlenbeck processes with autocorrelation times $1/\nu_i$. Very important and even crucial in applications is that this Maxwell-Langevin approach to viscoelastic 
subdiffusive dynamics allows for a straightforward Markovian embedding \cite{GoychukACP12}:
\begin{subequations}
\begin{eqnarray}\label{embedding1}
  &&\eta_0\dot{x}=f(x)-\sum_{i=1}^N k_i(x-x_i)+\sqrt{2k_BT\eta_0}\zeta_0(t), \\
  &&\eta_i\dot{x}_i=k_i(x-x_i)+\sqrt{2k_BT\eta_i}\zeta_i(t),
  \label{embedding2}
\end{eqnarray}
\end{subequations}
where $\zeta_i(t)$ are $N+1$ uncorrelated white Gaussian noises, $\langle\zeta_i(t)\zeta_j(t')\rangle=\delta_{ij}\delta(t-t')$, and $\eta_i=k_i/\nu_i$ are frictional coefficients of auxiliary Brownian particles. This Markovian dynamics in the space of $N+1$ dimensions can be propagated using well-established algorithms like stochastic Euler or stochastic Heun methods \cite{GardBook} without principal difficulties, with a well controlled numerical accuracy. By excluding the auxiliary variables $x_i$ in Eq. (\ref{embedding1}), (\ref{embedding2}) it is easy to show that the resulting GLE for the coordinate $x$ has indeed the memory kernel, which is presented by the sum of exponentials (\ref{Prony}), and the correlated noise term, which is the sum of corresponding Ornstein-Uhlenbeck processes. 
For this, one has to first rewrite (\ref{embedding2}) in terms of the viscoelastic force $u_i=k_i(x_i-x)$, and formally solve the resulting equation for $u_i$. This yields
\begin{eqnarray}\label{u_exact}
u_i(t)=-\int_0^t k_i e^{-\nu_i(t-t')}\dot x(t')dt' + \chi_i(t)
\end{eqnarray}
with 
\begin{eqnarray}\label{chi_exact}
\chi_i(t)&=&u_i(0)e^{-\nu_i t}\\&+&\sqrt{2k_BTk_i\nu_i}\int_0^t e^{-\nu_i(t-t')}\zeta_i(t')dt'. \nonumber
\end{eqnarray}
Each noise component $\chi_i(t)$ depends on $u_i(0)$, and all the noise components are mutually independent. Indeed, considering $u_i(0)$ as independent random Gaussian variables with $\langle u_i(0)\rangle=0$ and  $\langle u_i^2(0)\rangle=k_i k_BT$, one can show that $\chi_i(t)$  present wide sense stationary Gaussian stochastic processes with $\langle \chi_i(t)\chi_j(t')\rangle =k_BTk_i\delta_{ij}e^{-\nu_i|t-t'|}$. Substituting (\ref{u_exact}) in 
(\ref{embedding1}) establishes the stated equivalence \cite{GoychukPRE09,GoychukACP12}, 
provided  that the initial $x_i(0)$ in (\ref{embedding2}) are random Gaussian variables such that, $\langle x_i(0)\rangle =x(0)$, $\langle [x_i(0)-x(0)]^2\rangle=k_BT/k_i$. The considered Markovian embedding is exact, when the memory kernel is exactly the sum of exponentials (\ref{Prony}). 

The whole idea of Markovian embedding is very natural and sound in view of a dynamical origin of GLE dynamics: Instead of considering huge many thermal bath oscillators one replaces them by a handful of overdamped stochastic Brownian oscillators, with a nice physical interpretation in terms of a generalized Maxwell-Langevin theory of viscoelasticity \cite{GoychukACP12}. 
The  efficiency of the resulting numerical approach has a proven record \cite{GoychukPRE09,ChemPhys10,GoychukACP12,NJP12,MMNP13,PLoSONE14,PCCP14,PhysBio15}.
Upon a modification, this method can also be used for Markovian embedding of superdiffusive FLE dynamics \cite{SieglePRE10,SiegleEPL11}.
Clearly, it can be also considered as an independent approach to anomalous dynamics without any relation to FLE.

In any particular
case, one has to choose the embedding parameters appropriately, considering a trade-off between the numerical accuracy and feasibility of simulations on the required time scale. The accuracy is 
controlled by comparison with the exact results like one in  Eqs. (\ref{exact}), (\ref{exact2}).
In our simulations below we use for $\alpha=0.5$, $b=10$, $\eta_0=0.1$, $\nu_0=10^3$, $N=12$, $C_{0.5}(10)=1.3$, which warrants $3 - 5\%$ accuracy in numerics. 
The rms of potential $\sigma$ is fixed in simulations, whereas temperature varies in units of $\sigma/k_B$.
The time-step of integration was chosen $\Delta t=5\times 10^{-5}$, and the maximal time was $t_{\rm max}=2\times 10^5$. Stochastic Heun method (see in Appendix C) was used with double precision on high-performance graphical processors Tesla K20. In the ensemble trajectory simulations, $n=10^4$ particles were initially uniformly distributed within $[0,L]$ spatial interval with 10 different potential realizations in each case ($10^5$ particles in the ensemble averaging). 
Random potentials were generated as described above, in accordance with \cite{SimonFNL}, and simulations were run with periodic boundary conditions. It took typically about 5 days of computational time for each ensemble-averaged curve presented below. The results were first tested against the exact analytical result in Eq. (\ref{exact2}) in the absence of random potential. The numerical results coincide in this case with the analytical result within the width of the plotted curves like in Fig. 2, a in Ref. \cite{GoychukPRE09}, and especially, 
Figs. 5, 6 in Ref. \cite{GoychukACP12}, Fig. 2 of \cite{MMNP13} and inset of Fig. 1 in Ref. \cite{PRE13}.  \\

\section{Results and Discussion}

\subsection{Ensemble averaging}

\begin{figure}[h]
\centering
\resizebox{0.8\columnwidth}{!}{\includegraphics{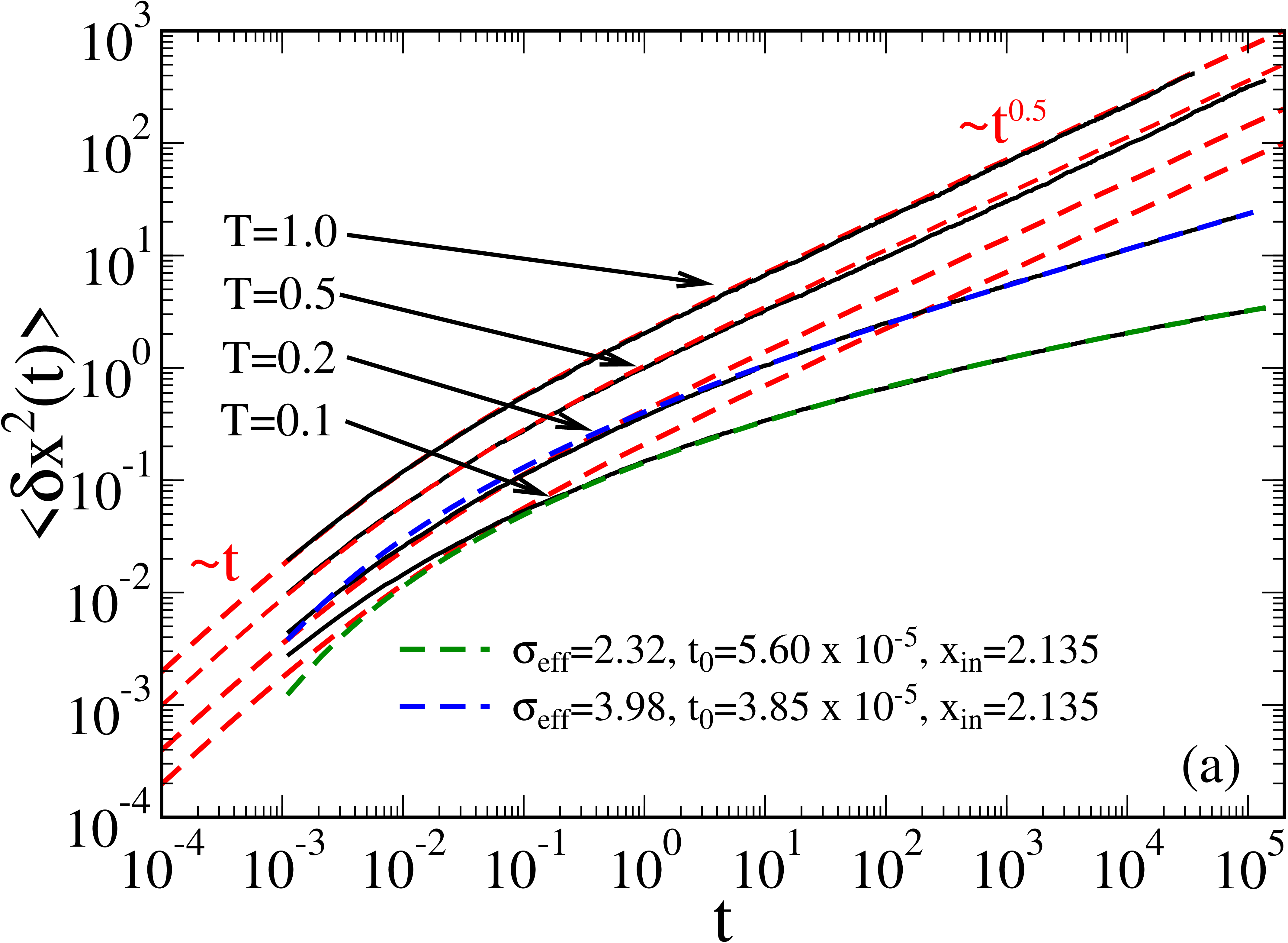}}\\[1cm]
\resizebox{0.8\columnwidth}{!}{\includegraphics{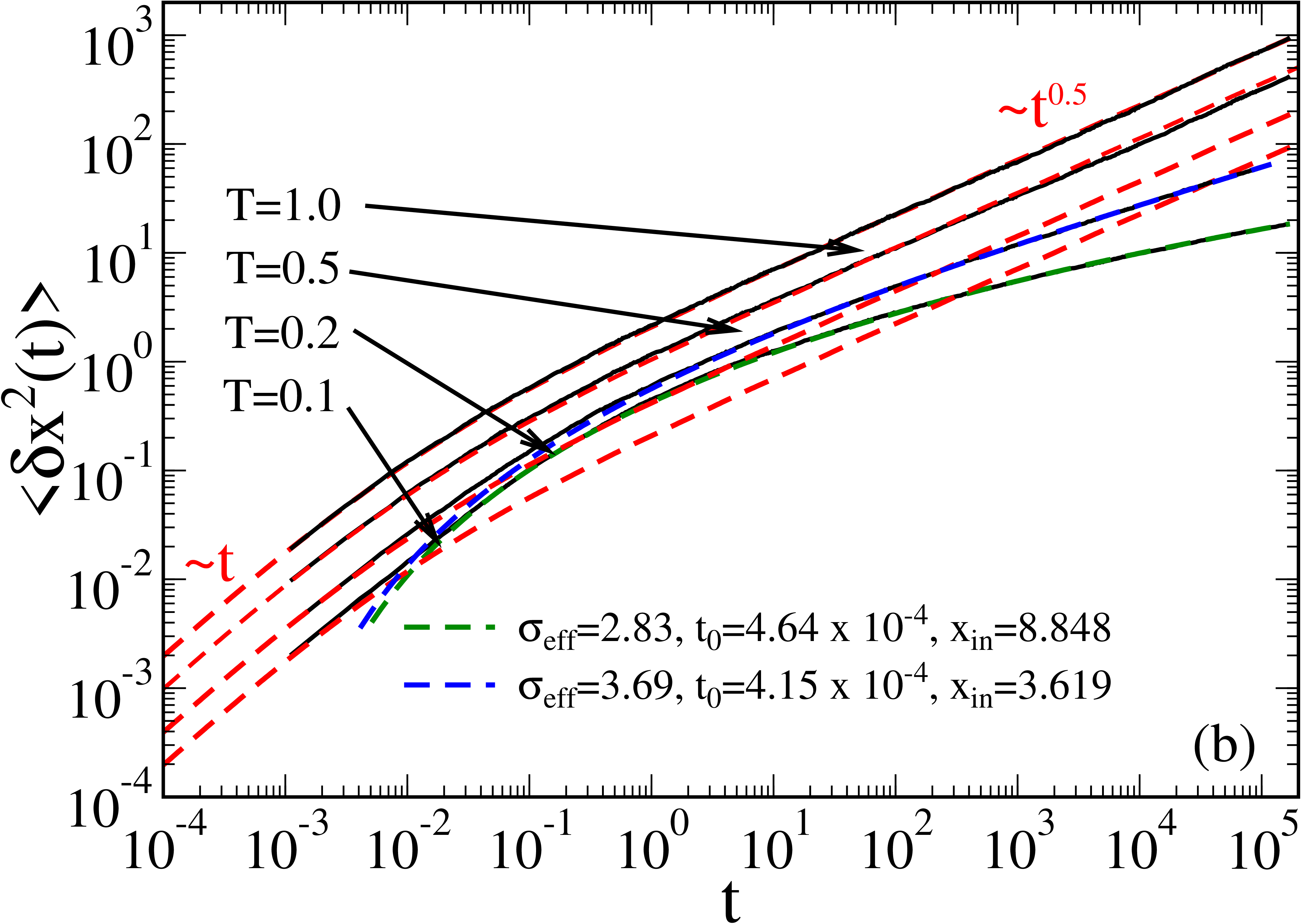}}
\caption{(Color online) Ensemble-averaged mean squared displacement versus time in units of $\tau_r=(\lambda^2\eta_\alpha/\sigma)^{1/\alpha}$ for
	different values of $k_BT$ in units of the disorder strength $\sigma$ for  (a) 
 exponential decay of correlations and (b) power-law decay with $\gamma=0.8$. The fit of 
 the numerical results (full black lines) is performed for $T=0.2$ with the expression (\ref{central}) and for $T=0.1$ using (\ref{Sinai}). The fitting parameters are shown in the plot. Dashed red lines depict exact results for free subdiffusion in accordance with Eq. (\ref{exact2}): $\alpha=0.5$, $\eta_0=0.1$ and $\tau_0=0.01$.
}
\label{Fig2}       
\end{figure}

\begin{figure}[h]
\centering
	\resizebox{0.8\columnwidth}{!}{\includegraphics{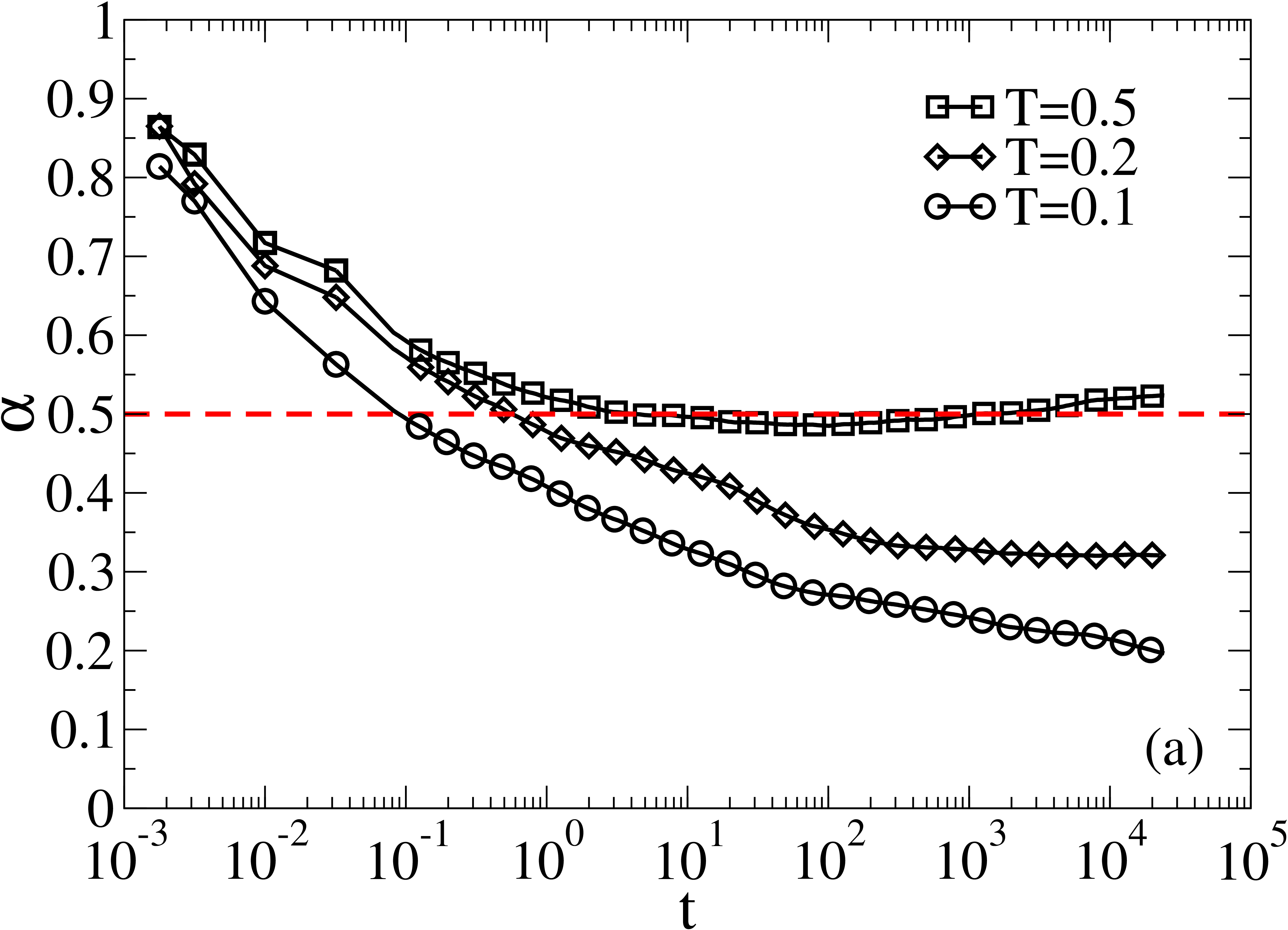}}\\[0.8cm]
	\resizebox{0.8\columnwidth}{!}{\includegraphics{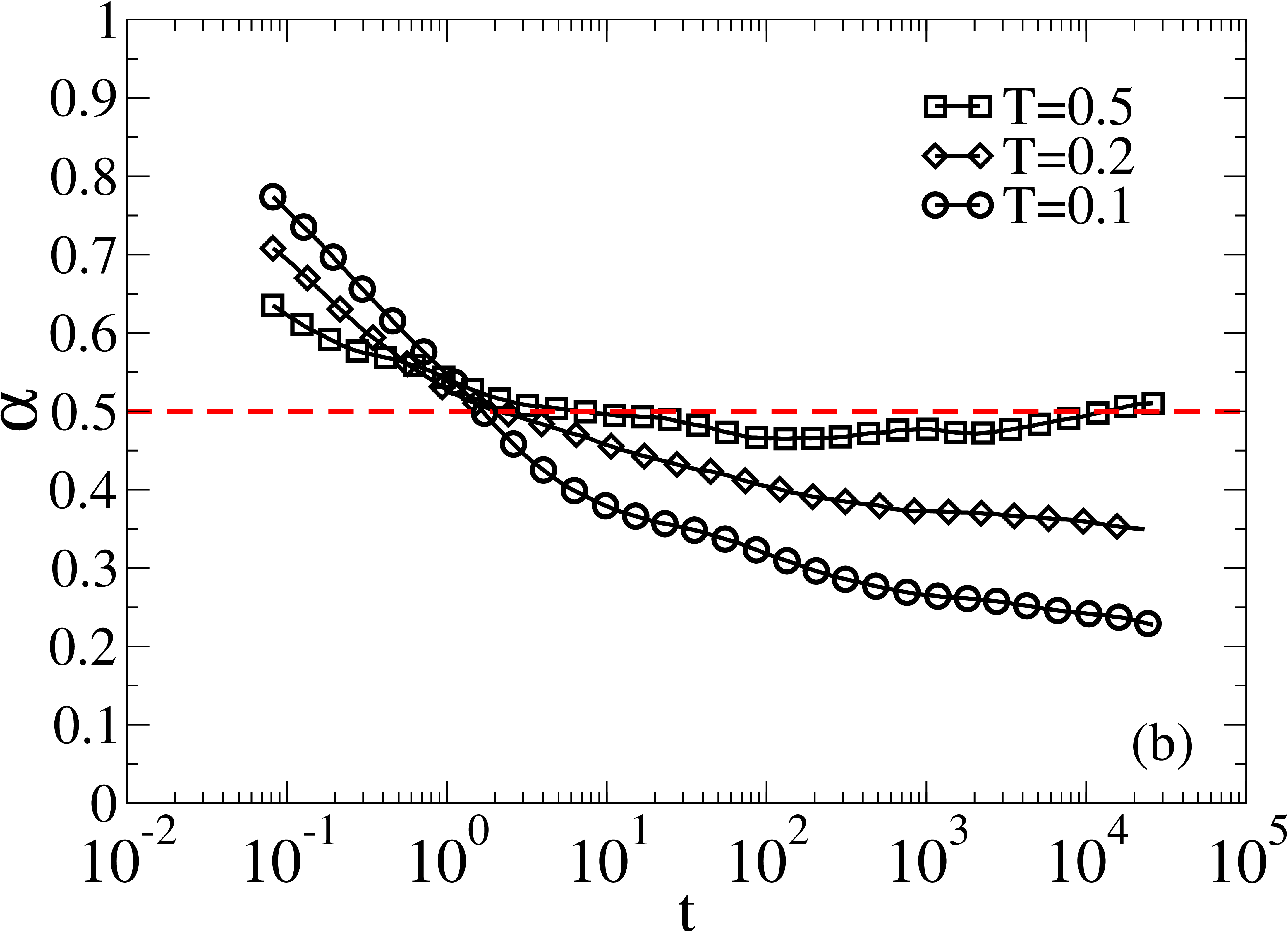}}
	\caption{(Color online)  Time-dependent power law exponent $\alpha(t)$
		for an assumed subdiffusive law $\langle\delta x^2(t)\rangle\propto  t^{\alpha(t)}$
		obtained as the logarithmic derivative of the traces in Fig.~\ref{Fig2}, for
		different temperatures in the case of (a) exponential correlations, (b) power law
		correlations with $\gamma=0.8$. }
	\label{Fig3}       
\end{figure}

We first concentrate on the ensemble averaging. The results are shown in Fig. \ref{Fig2} for the exponentially decaying correlations in part (a) and for the power-law decaying correlations with $\gamma=0.8$ in part (b), for several different values of temperature starting from $T=1$ and ending with $T=0.1$. The first striking feature for both types of correlations is that random potential practically does not matter for $T=1$ and the results are not different from the exact result of potential-free subdiffusion depicted by a broken red line in accordance with Eq. (\ref{exact2}). This is not a trivial feature at all, even if it could be expected from the earlier results on viscoelastic subdiffusion in periodic potentials \cite{GoychukPRE09,GoychukACP12}. However, intuition says that a combination of slowness caused viscoelastic effects with sluggishness  caused by random potential should result into an ultraslow behavior. This intuition is wrong. Very differently from memoryless diffusion in stationary Gaussian potentials, which is asymptotically normal and exponentially suppressed by disorder, viscoelastic subdiffusion is not suppressed  asymptotically by disorder at all, on the ensemble level.  
This result is very surprising indeed because another fractional dynamics in such random potentials, namely fractional Fokker-Planck dynamics, predicts a very different result, see in Appendix A,
$\langle \delta x^2(t)\rangle=2D_\alpha \exp[-\sigma^2/(k_BT)^2] t^\alpha/\Gamma(1+\alpha)$, i.e. the renormalization factor is the same for as for normal diffusion. 
Especially in the case exponential correlations this result of viscoelastic fractional subdiffusion is very surprising even for $T=1$, as soon one realizes that in this case the amplitude of potential fluctuations can largely exceed 
$k_BT$ well within a distance of the correlation length, see in Fig. \ref{Fig1}. However, with the lowering  temperature the influence of potential becomes visible already for $T=0.5$, although the potential-free asymptotics is almost reached at the end point of simulations in Fig. \ref{Fig2}, and the influence is really small, barely detectable. For $T=0.1$, it becomes very distinct, and the corresponding transient regime lasts indeed very long: No slightest signature of an asymptotic regime is even present  in Fig. \ref{Fig2} for $T=0.1$. The corresponding asymptotics is simply impossible to reach numerically.  Instead of a power-law subdiffusion, one clearly detects a nominally ultra-slow logarithmic diffusion of the Sinai type:
\begin{eqnarray}
\label{Sinai}
\langle \delta x^2(t) \rangle \approx 
 x_{\rm in}^2 
\left [(k_B T/\sigma_{\rm eff})\ln(t/t_0) \right ]^4
\end{eqnarray}
with three fitting parameters: $\sigma_{\rm eff}$, $ x_{\rm in}^2 $, and $t_0$, two of which can be combined in the only one, $x_{\rm in}^2/\sigma_{\rm eff}^4$, in this case. It describes numerics nicely over about $7$ time decades for both types of correlations. For a larger $T=0.2$, numerics are fitted well by a more complex, yet only three parameters dependence \cite{Goychuk2017}
\begin{eqnarray}
\label{central}
\langle \delta x^2(t) \rangle = x_{\rm in} ^2 
\left\{e^{[(k_BT/\sigma_{\rm eff})\ln (t/t_0)]^2}-1 \right\}^2\;.
\end{eqnarray}
It follows from a scaling consideration assuming that a typical time to travel a certain distance $x$ is defined in an Arrhenius manner by the largest potential barrier met on the pathway and the fact that this largest barrier scales as \cite{ZhangPRL,HanesJPCM,Goychuk2017} 
\begin{eqnarray}\label{second}
\delta U_{\rm max}\sim 2\sigma \sqrt{2 \ln(x/x_{\rm in})}  
\end{eqnarray}
with the distance. Indeed, let us estimate a typical time $t$ it takes for a particle to travel the distance $x$ starting at $x_0$. It is reasonable to assume that on the intermediate time scales, where the presence of potential is very essential, this time is defined, like in the case of normal diffusion,  by the largest barrier met on the particle's way,
$t=t_0\exp[|\delta U_{\rm max}(x)|/(k_BT)]$, where $t_0$ is a prefactor. From this scaling ansatz, upon taking (\ref{second}) into account, we obtain the estimate in (\ref{central}) with $\sigma_{\rm eff}=2\sqrt{2}\sigma$. Given a very crude character of this estimate, $\sigma_{\rm eff}$ should be considered as a fitting parameter. Like for memoryless diffusion, this result
 holds also for viscoelastic subdiffusion because a typical mean time to overcome a potential barrier does scale in Arrhenius manner with its height \cite{GoychukPRE09,GoychukACP12}. Namely this kind of behavior dominates in the transient regime, where the influence of potential on viscoelastic subdiffusion is very essential. Sinai diffusion in Eq. (\ref{Sinai}) just follows from Eq. (\ref{central}) as the lowest order expansion in $k_BT/\sigma_{\rm eff}$.
The fitted values of $\sigma_{\rm eff}$  agree actually fairly well with the theoretical value 
$\sigma_{\rm eff}=2\sqrt{2}\sigma\approx 2.83\sigma$ \cite{Goychuk2017}.
The agreement of the fitted values with theoretical value of $x_{\rm in}=\pi\lambda /\sqrt{\gamma}\approx 3.51 \lambda$ for $\gamma=0.8$ \cite{Goychuk2017} is also rather good for power law correlated potentials, see especially for $T=0.2$.
For singular model with exponential correlations, which predicts $x_{\rm in}=\pi\sqrt{\lambda \Delta x/2}\approx 0.315 \sqrt{\lambda}$ for $\Delta x=0.02$, the agreement worsens. Nevertheless, scaling argumentation of Refs. \cite{Bouchaud1990,Goychuk2017} works surprisingly good, given its very rough character, also for viscoelastic subdiffusion in random potentials. Notice that in power-law correlated potentials, Sinai-like subdiffusion is essentially faster in absolute terms than one in exponentially correlated potentials. The reason becomes immediately clear from Fig. \ref{Fig1}. This is because power-law correlated disorder is much smoother, and the maximal barrier met on the same distance is essentially smaller than in the case of exponential correlations.
 Furthermore, an interesting transient effect on the ensemble level is that viscoelastic subdiffusion in random potential can be even faster than the potential-free subdiffusion, see for power law correlations and $T=0.1$ in Fig. \ref{Fig2}, b. This is because of an alternating local bias felt by each separate particle \cite{Goychuk2017}. Such a local bias and the resulting random drift are responsible e.g. for the Golosov phenomenon in the case of genuine Sinai diffusion \cite{Golosov84,Bouchaud1990}. Golosov phenomenon describes at the first look paradoxical effect that in an environment with random bias two particles starting nearby do not diffuse strongly apart being subjected to one and the same local bias in the environment \cite{Golosov84,Bouchaud1990}. The observed phenomenon is different. However, it has precisely the same physical origin: a strong local bias which alternates randomly its direction.  Single-trajectory averages, see below, do not show such a paradoxical feature. Then, subdiffusion is always suppressed by random potential.

\begin{figure*}[]
\centering
	\includegraphics[width=6cm]{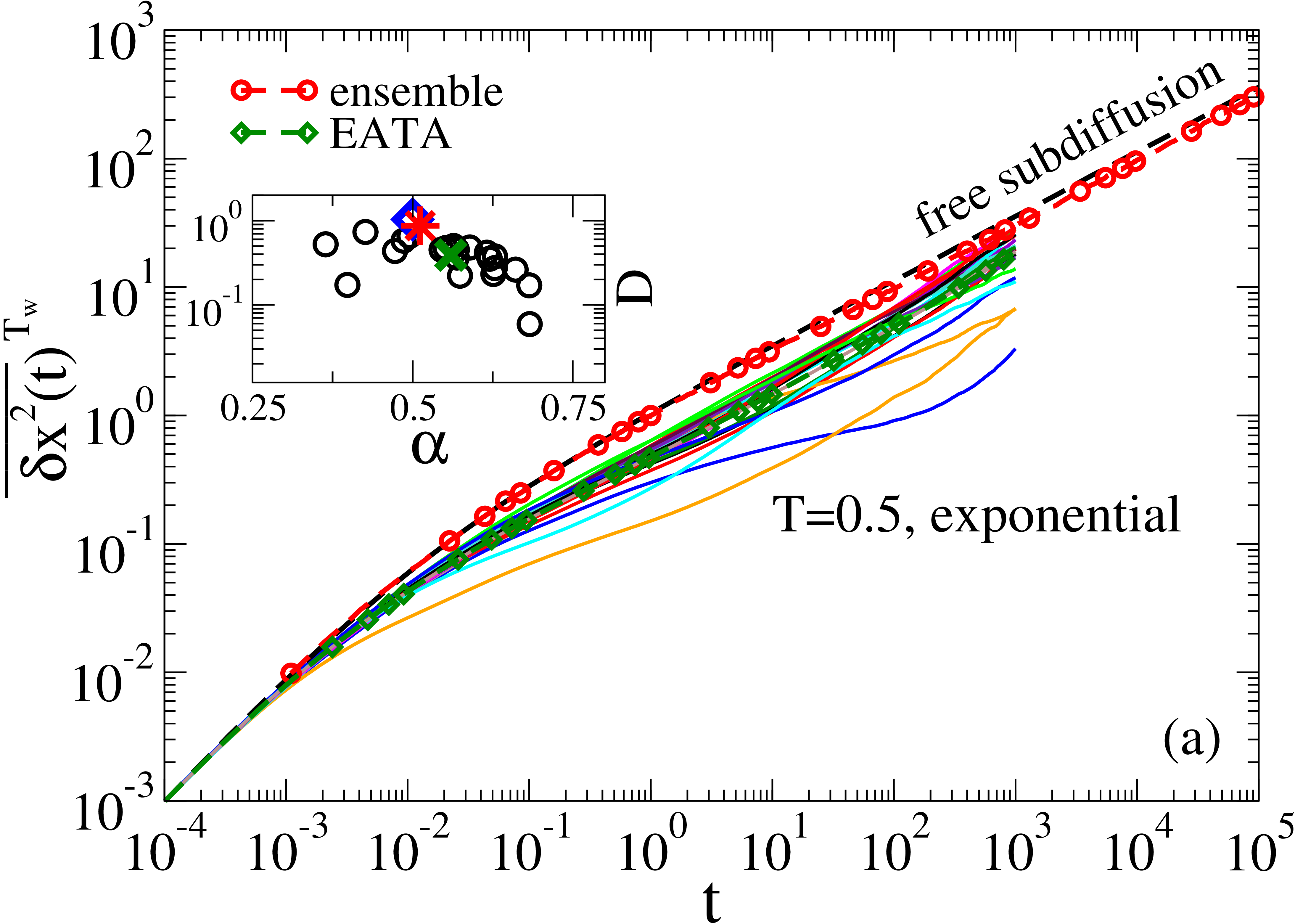} 
	\hspace*{0.8cm}
	\includegraphics[width=6cm]{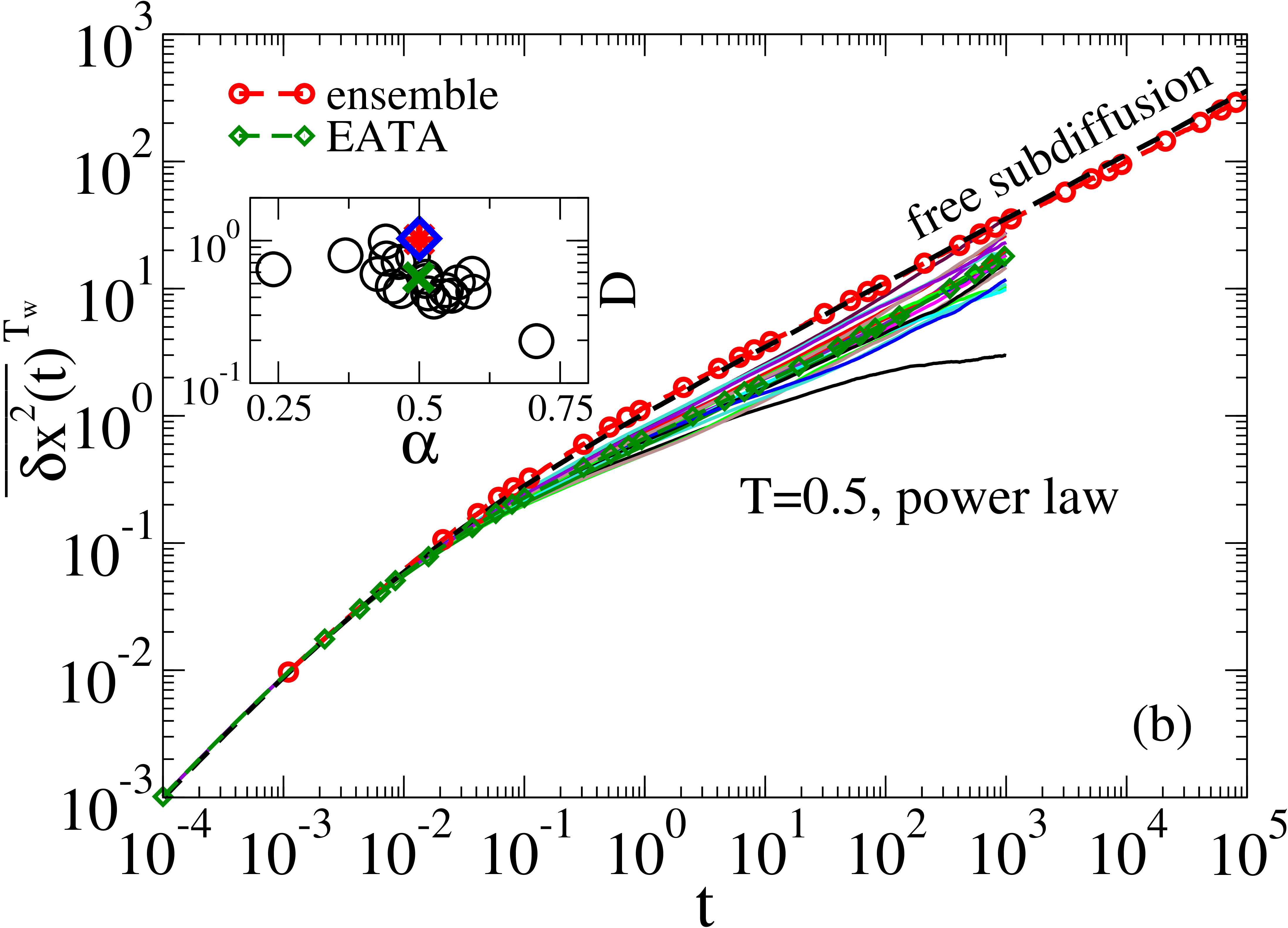} \\ [0.8cm]
	\includegraphics[width=6cm]{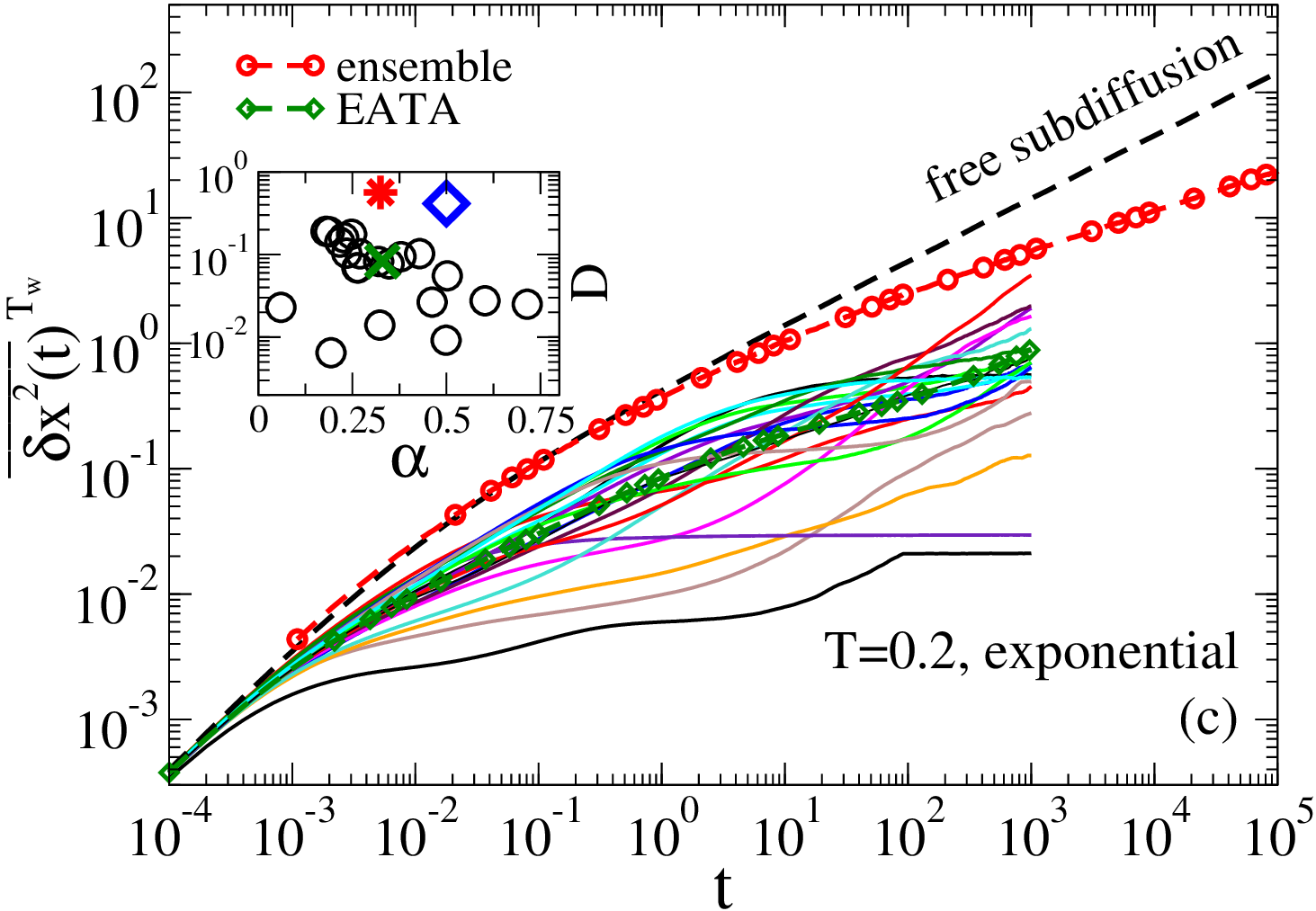} 
 	\hspace*{0.8cm}
	\includegraphics[width=6cm]{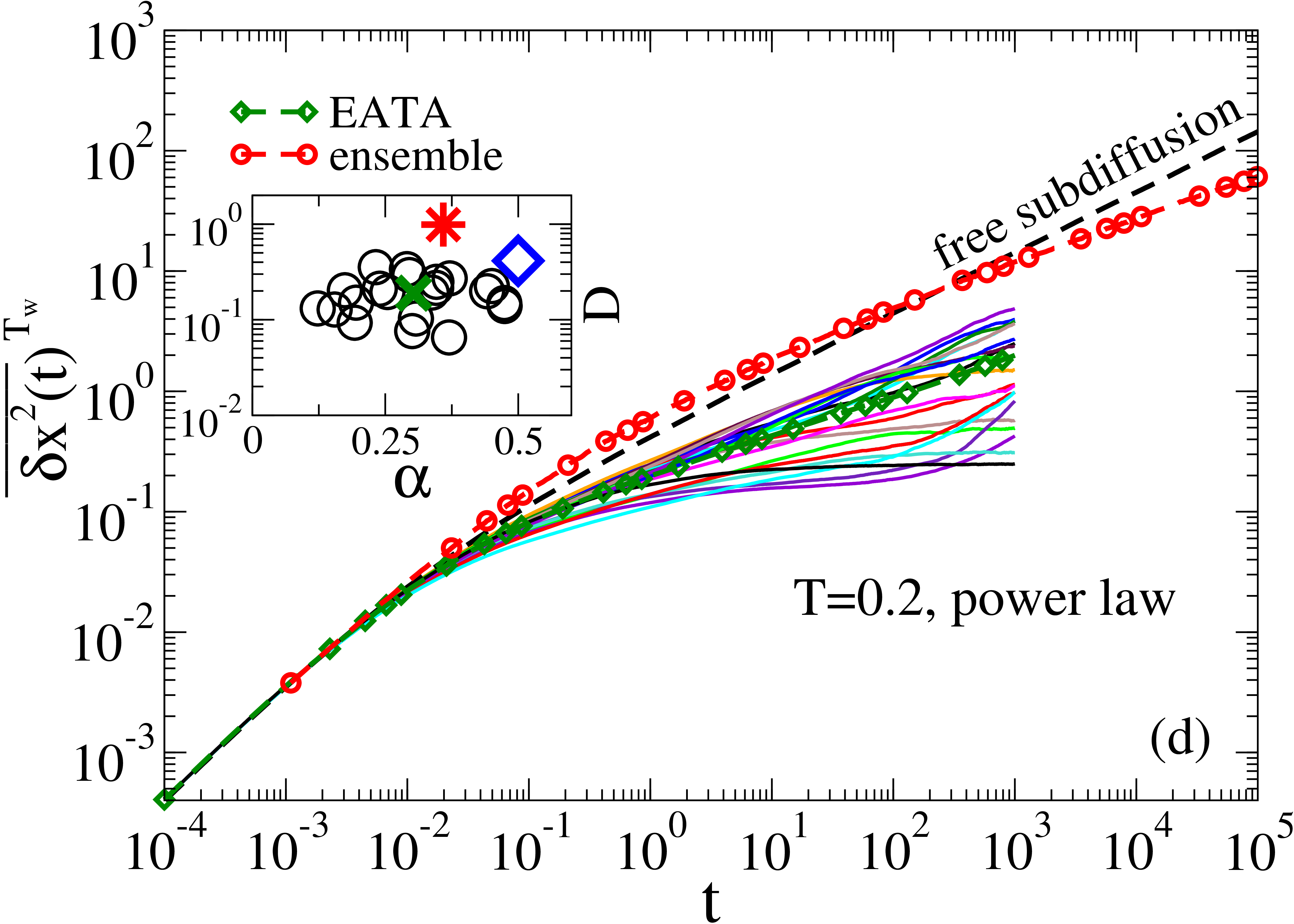}
	\caption{(Color online) Single-trajectory, time-averaged mean squared displacement
		for two values of temperature $T=0.5$ and $T=0.2$, and two types of correlations shown in each panel. 
		The trajectories time length was ${\cal T}_{\rm w}= 10^5$.
		20 trajectory averages were made for particles starting from different locations. 
		They are depicted with solid lines.
		The results of the ensemble-averaged, as well as ensemble-averaged time-averaged (EATA), and
		potential-free subdiffusion are also depicted for comparison.
Insets show the distribution of the scaled subdiffusion coefficient $D$ and power law exponent $\alpha$ for single-trajectory fits with dependence $Dt^\alpha$. The green cross therein corresponds to the averaged values of $D$ and $\alpha$, while the red star to the ensemble-averaged result. The result of free subdiffusion is depicted for comparison as a blue diamond in each inset. 
		 Remarkably, the ensemble average is only slightly suppressed by the random potential even for $T=0.2$ in the case of power law correlations at the end point of simulations, see in the panel d, whereas transiently it is even faster. However, single-trajectory averages are suppressed essentially stronger. Generally, scatter in single-trajectory averages is visibly stronger for exponential correlations. }  
	\label{Fig4}
\end{figure*}

An alternative to Eq. (\ref{central}) way to represent the results is to introduce a time-dependent exponent $\alpha(t)$ of power-law subdiffusion
\begin{eqnarray}
\label{intermed}
\langle \delta x^2(t)\rangle\approx x_{\rm in}^2 [t/t_0]^{\alpha(t)}\;.
\end{eqnarray}
Its behavior is depicted Fig. \ref{Fig3}. For Sinai-like diffusion at $T=0.1$ and $T=0.2$, $\alpha(t)$ declines in time. It should reach a minimum \cite{Goychuk2017} and then logarithmically slow grow, $\alpha(t)\propto \log(t)$, as Eq. (\ref{central}) predicts, \cite{Goychuk2017} for a possibly very long period of intermediate times, until the assumptions which lead to (\ref{central}) remain valid.
Indeed, in the course of time, when the unity becomes negligible in Eq. (\ref{central}), it reduces to Eq. (\ref{intermed}) with 
\begin{eqnarray}\label{intermed2}
\alpha(t)=2(k_B T/\sigma_{\rm eff})^2\ln(t/t_0).
\end{eqnarray}
For $T=0.2$ and exponential correlations, 
the minimum is indeed reached at $\alpha_{\rm min }\approx 0.32$ in Fig. \ref{Fig3}, a, which is approximately the same value as for normal diffusion in this potential \cite{Goychuk2017}.
However, it is still not achieved in Fig. \ref{Fig3}, b, for $T=0.2$ in the case of power law correlations. Furthermore, for $T=0.1$, it is not achieved for the both types of correlations. Unfortunately,
the regime of logarithmically growing $\alpha(t)$ is 
numerically not achievable in our simulations even for exponential correlations and $T=0.2$. To find it, one should probably propagate the dynamics by a factor of 100 longer. This is clearly not feasible computationally at present. 
 This behavior is in contrast with one of the major features of memoryless diffusion in the studied random potentials \cite{Goychuk2017}, where such a long-lasting  intermediate regime was clearly detected, for both exponential and power-law potential correlations. This is because in the case viscoelastic subdiffusion various transient regimes last much longer. However, computationally it is much more demanding and the corresponding time scale is difficult to reach.

\subsection{Single-trajectory averages}

Single-trajectory averages \cite{HePRL08,LubelskiPRL08,Barkai,MetzlerPCCP}
 \begin{equation}\label{single}
 \overline{\delta x^2(t)}^{{\cal T}_w}=\frac{1}{{\cal T}_w-t}\int_0^{{\cal T}_w-t}\Big[ \delta x(t|t')\Big]^2dt'
 \end{equation}
 of the mean-squared displacement $\delta x(t|t')=x(t+t')-x(t')$ over the maximal time
 window ${\cal T}_w$  present a great interest, especially for experimentalists who often simply do not have a possibility to deal with macroscopically many particles. To avoid a trivial statistical scatter in Eq. (\ref{single}), the maximal time $t$ should be much smaller than  ${\cal T}_w$. In the numerical results depicted in Fig. \ref{Fig4} it is just 1\%. Remarkably, even for $T=0.5$ the scatter in single-trajectory averages is strong. It is clearly more pronounced in the case of exponential correlations for the reason, which is already obvious. Interesting, fitting the single-trajectory averages as $\overline{\delta x^2(t)}^{{\cal T}_w}=Dt^\alpha$, with a trajectory-specific scaled anomalous diffusion coefficient $D$ and corresponding power exponent $\alpha$, gives the corresponding values broadly scattered, see in insets in Fig.~\ref{Fig4}. The mean value $\bar \alpha$ of this trajectory-specific $\alpha$ is depicted by a green cross in the corresponding insets. For power law correlations (part \textbf{b}), $\bar \alpha \approx 0.50$ is the the same as for the ensemble-averaged curve (red star) and free subdiffusion (blue diamond), while for the exponential correlations in the part \textbf{a} it is slightly different,  $\bar \alpha \approx 0.56$. This is an interesting feature. For example, in Ref. \cite{Golding} single-trajectory averages for subdifusing mRNA molecules are scattered around the same $\alpha=0.70$ (different from $\alpha=0.5$ fixed in our numerics here). Moreover, in Ref. \cite{MagdziarzPRL} it has been shown that the data in Ref. \cite{Golding}
 are more consistent with a fractional Brownian motion than a CTRW subdiffusion. Indeed, we see in the inset of our Fig. \ref{Fig4}, a that $D$ is scattered over a range of about 40 between the minimal and maximal values, whereas in experiment the scattering range is about 100. Taking into account that the size of mRNA is also distributed \cite{Golding} and the results of Ref. \cite{MagdziarzPRL}, it is indeed looks likely that 
viscoelastic subdiffusion in a random environment is better suited to explain subdiffusion in bacterial cytoplasm than CTRW.
 
 For a stronger disorder of  $\sigma=5\;{k_BT}$ in Fig. \ref{Fig4}, c,d, $\alpha$ is scattered stronger
 and its mean value is smaller, $\bar \alpha=0.33$ in the part \textbf{c} and $\bar \alpha=0.303$ in the part \textbf{d}, which correlate the corresponding fitting values of the ensemble-averaged subdiffusion, 
 $0.325$ and $0.36$, correspondingly, see also
 the corresponding end points in Fig. \ref{Fig3}. These values are not related to $\alpha$ of free subdiffusion and have a very different origin, the same as for normal diffusion in such potentials \cite{Goychuk2017}, see also above. The scatter of $D$ becomes also more pronounced.
  
 Notice also that while the ensemble-average of single-trajectory averages, EATA, in figure \ref{Fig4} gradually converges to the ensemble-averaged result, in the case $T=0.5$, some of the single-trajectory averages can look yet very different. In this respect, one should mention that many experimental data on subdiffusion in living cells seem to clearly point out on the viscoelastic mechanism of this subdiffusion upon use of 
 several strict criteria \cite{Robert,SzymanskiPRL,WeissPRE13,MagdziarzPRL}. However, other researchers doubt it because single trajectory averages reveal essential non-ergodic features \cite{WeigelPNAS,TabeiPNAS}.  A tentative resolution of this paradox is that the discussed biologically related anomalous diffusion is viscoelastic subdiffusion in a random, inhomogeneous and fluctuating environment. It seems almost obvious, on physical grounds. Differently from Ref. \cite{TabeiPNAS}, we model this fact by imposing a random potential on viscoelastic subdiffusion, rather than subordinating physical time to a random clock of CTRW. Indeed a typical mesh size of random actin meshwork in eukaryotes cytosol and model polymeric fluids is $0.1 - 1$ $\mu$m depending on the actin concentration \cite{WongPRL}. Let us take it to be  
$\lambda \approx  0.308$ $\mu$m and associate it with the correlation length of random potential. Furthermore, let us consider diffusion of globular proteins of the radius $R=2.5$ nm (a typical one) in such a system. Actin meshwork is charged and globular proteins are also typically charged 
\cite{WongPollack10,Grosberg02,Messina09}. This will cause a screened (by mobile ions) electrostatic interaction. The strength can be variable depending on the mesh size, the screening length, and the size of particle. The whole problem is highly nontrivial and given the complexity of electrostatic interactions in soft matter  \cite{WongPollack10,Grosberg02,Messina09} 
it does not seem to be even properly approached at the moment.  
 Nevertheless, given a typical strength of electrostatic interactions in soft matter  it is not unreasonable to take $\sigma=(2 - 5)\;k_BT$ as a first reasonable guess in our estimate.
 Indeed, distribution of binding energy of regulatory proteins to DNA tracks has also the same 
 typical range \cite{LassigReview,GerlandPNAS,BenichouPRL09}.
 The subdiffusion coefficient of a particle of radius 2.5 nm in cytosol should be about the same as for a gold nanoparticle of the same radius in HeLa cells in Ref. \cite{Guigas}. It is estimated to be $D_\alpha\approx  0.644 \;{\rm \mu m^2/s^{1/2}} $ in our notations (see Table I in Ref. \cite{PRE12b}). The inertial effects in such a case are completely negligible and the time scale parameter $\tau_r$ is estimated to be  $\tau_r\approx 5.48 $ ms, see in Appendix B. Hence, the maximal time in our Fig. \ref{Fig4} is about $5.48$ s for single trajectory averages. In accordance with our results,  for $\sigma=2\;k_BT$ they would be broadly scattered on this time scale as in Fig. \ref{Fig4}, a, b, even if the averaging time window ${\cal T}_w$ would be 548 s long. However, the ensemble average would be almost independent of the presence of random potential, like in
Fig. \ref{Fig2} for $T=0.5$. This is a first crude idea to resolve some current controversies.
However, a further quantitative analysis of the available experimental data from the discussed perspective of viscoelastic subdiffusion in random potentials  is required and welcome.

\subsection{Escape time distribution}

\begin{figure}
\centering
\resizebox{0.75\columnwidth}{!}{\includegraphics{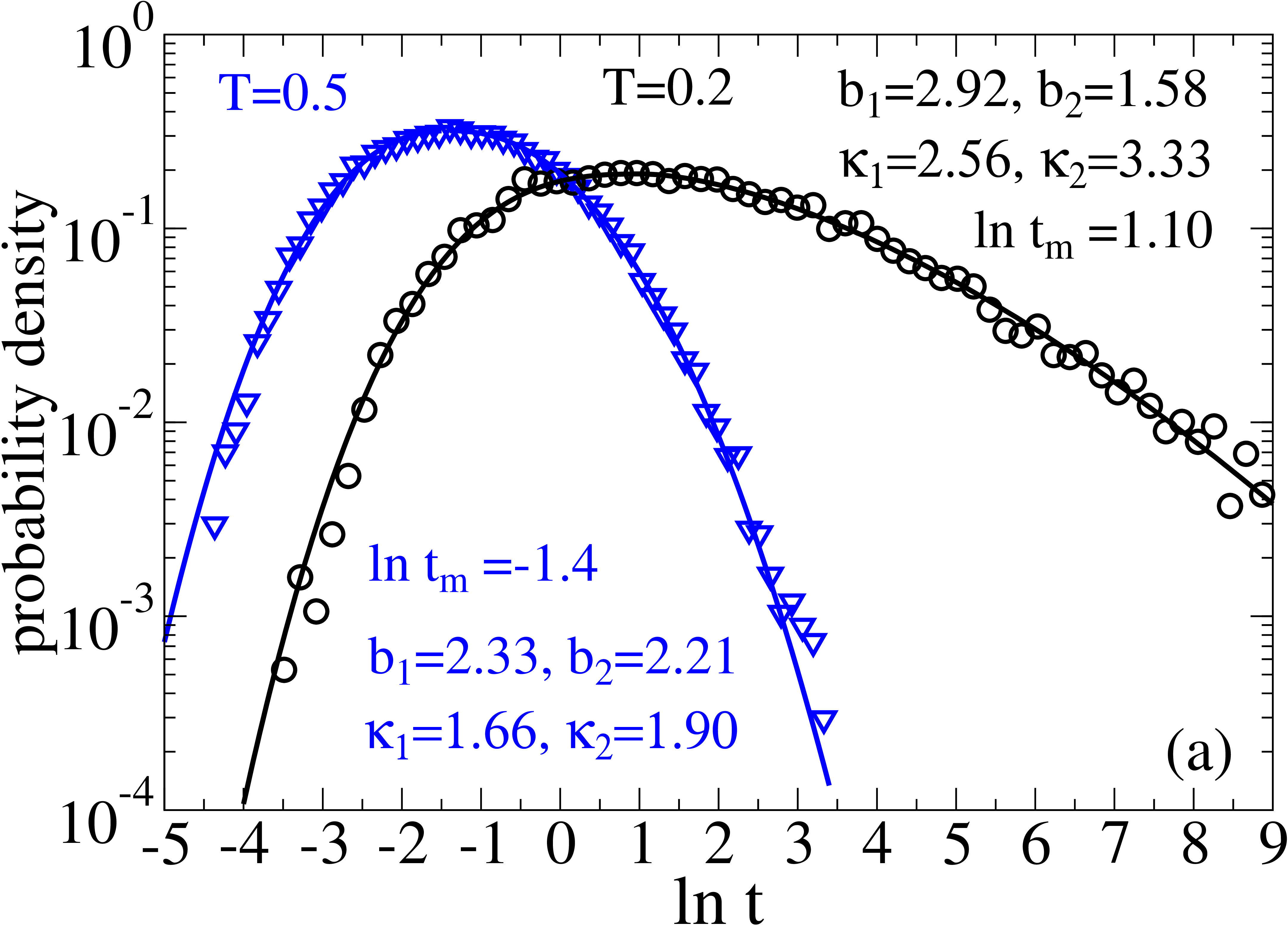}}\\[0.8cm]
\resizebox{0.75\columnwidth}{!}{\includegraphics{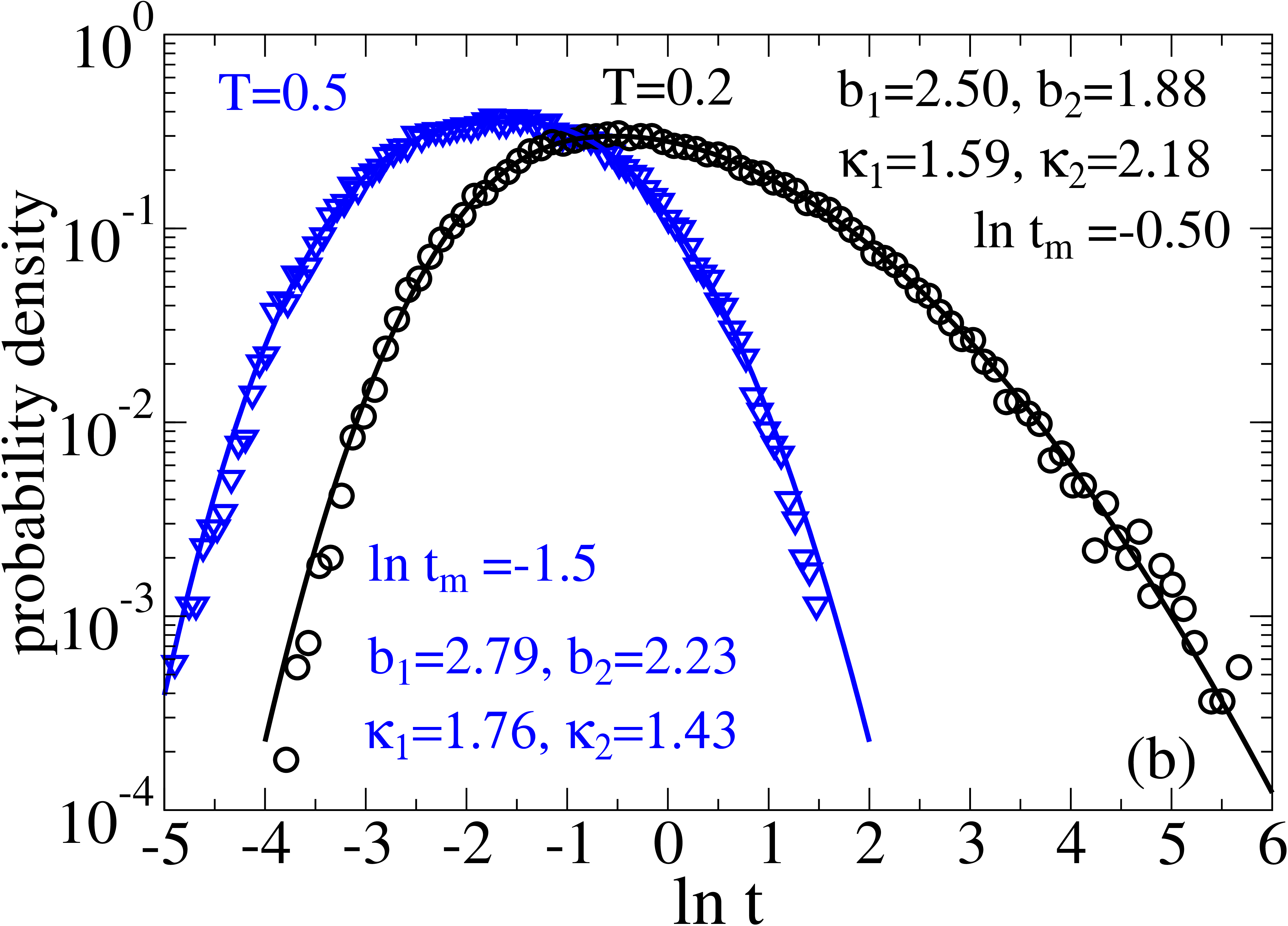}} \\[0.8cm]
\resizebox{0.75\columnwidth}{!}{\includegraphics{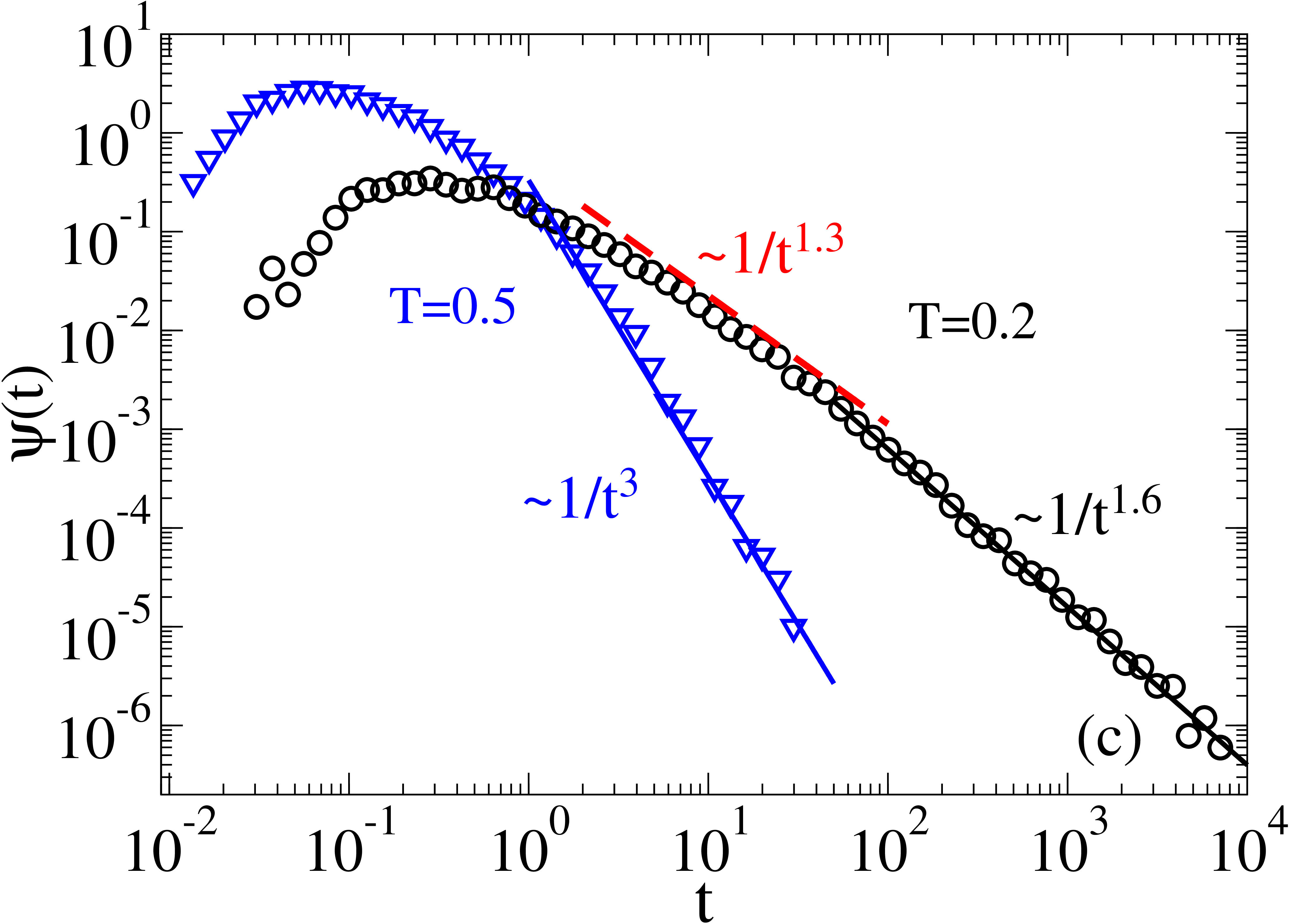}}  
\caption{(Color online)  Probability
density function of log-transformed first escape times, $z=\ln t$, from the interval $[-\lambda,\lambda]$ for
two temperatures $T=0.5$ (in blue) and $T=0.2$ (in black). The cases in the different panels are:
(a) exponential correlations, (b) power-law correlations with $\gamma=0.8$. The symbols represent
simulations data, the lines correspond to a fit with the probability density (\ref{distro}).
Parameters are shown in the panels. With decreasing temperature the distributions become broader
and the parameter $b_2$ smaller. In (c), probability densities of the original non-transformed time variable are plotted which correspond to the part (a). In this case,
$\psi(t)$ seems to show some parts with power law dependencies. Especially confusing is the case of
$T=0.2$, where the part of distribution with negative exponent $-1.3$ (indicated in red), in a conjunction with
$\alpha(t)\approx \alpha_{\rm min}\approx 0.32 $ in Fig. \ref{Fig3}, a and broad scatter of
single trajectory averages in Fig. \ref{Fig4}, c can be erroneously interpreted within a CTRW theory with divergent mean residence time. } 
\label{Fig5}        
\end{figure}

The success of the scaling argumentation extended from normal to subdiffusive viscoelastic dynamics in stationary Gaussian potentials suggests that the escape time distributions should also be similar. We consider escape of the particles out of $[-\lambda,\lambda]$ spatial interval, which are initially located at its center. The distribution of logarithmically transformed escape times, $z=\ln t$, is plotted in Fig. \ref{Fig5}. Indeed, a generalized log-normal distribution of Ref. \cite{Goychuk2017}
\begin{eqnarray}
\nonumber
\psi(t)&=&\frac{C}{t}\Big[ e^{-|\ln(t/t_m)/\kappa_1|^{b_1}}\theta(t_m-t)\\
&&+e^{-|\ln(t/t_m)/\kappa_2|^{b_2}}\theta(t-t_m)\Big],
\label{distro}
\end{eqnarray}
where $C=b_1b_2/[b_2\kappa_1\Gamma(1/b_1)+b_1\kappa_2\Gamma(1/b_2)]$ is a
normalization constant, $b_{1,2}>1$, and $\kappa_{1,2}>0$, fits excellently to the numerical data for the both considered models of correlations. General features are similar to those of memoryless diffusion. The escape  density has a maximum at $\ln t_m$ value of the logarithmically transformed time variable.  Furthermore, escape in the power-law correlated potentials occurs much faster than in the exponentially correlated potentials. The exponent $b_2$ is strongly temperature dependent. With lowering temperature it becomes smaller and closer to one. However, all the moments of RTD remain finite.
Notice that this generalized log-normal distribution can sometimes be easily mistaken for a power law, if to plot it in doubly logarithmic coordinates for the original non-transformed  time variable as e.g. in Fig. \ref{Fig5}, c, in the case of exponential correlations. 
It is also reminiscent of a stretched exponential distribution \cite{GoychukPRE09}.
This is why to know the moments of experimental distributions is important, as well as using other representations of experimental data, upon a transformation of random variable, like in our Fig. \ref{Fig5}, a, b. This would allow a new look on the existing experimental data, such as e.g. in Ref. \cite{WongPRL}, in the light of our model. Indeed, it is very tempting to interpret 
$\psi(t)\propto 1/t^{1.3}$ in Fig. \ref{Fig5}, c for $T=0.2$ in conjunction 
with $\alpha(t)\approx \alpha_{\rm min}\approx 0.32 $ in Fig. \ref{Fig3}, a, and a strong scatter of single-trajectory averages in Fig. \ref{Fig4}, c within the traditional CTRW theory with divergent mean residence time, like done in Ref. \cite{WongPRL}.  However, real physics in our particular case is very different. In our work, these results are produced by viscoelastic subdiffusion in a random Gaussian potential.
Unlike the case of CTRW subdiffusion \cite{MetzlerPCCP},  single-trajectory averages have in the studied case of viscoelastic subdiffusion an averaged $\bar \alpha$, which correlates well with the power law exponent of the ensemble-average, see in  Fig. \ref{Fig4}. This can provide an important experimental criterion to distinguish among various theoretical explanations possible.

\subsubsection{Digression on dimensionality}

The question of whether one can or not directly apply the results obtained within a one-dimensional model to two- or three-dimensional diffusion in living cells, or dense heterogeneous polymeric liquids is not trivial. First of all, in the one-dimensional case the particle cannot avoid a trap, or a barrier on its way. However, in 2d and 3d it can find the ways around. This remark is especially important for viscoelastic subdiffusion, which can thoroughly explore the space. Indeed, the fractal Hausdorff dimension of fBm
  trajectories occupying 3d space is \cite{Feder} $d_H = 2/\alpha$, for $2/3 < \alpha <
  1$, and $d_H = 3$, for $0 < \alpha < 2/3$. Hence, for $\alpha<2/3$ the
  fBm fills densely the three-dimensional Euclidean space and it can find all possible ways around.
Thus, our first expectation is that in 2d and 3d viscoelastic subdiffusion will generally more easily overcome the medium's disorder than in 1d. At the same time, it will show a significant scatter in single-trajectory averages. This is namely a kind of behavior which is often observed, and which our 1d theory predicts. To reveal the regime of Sinai like diffusion in higher dimensions seems less likely, unless the disorder is very irregular, of singular type, like in the case of exponential correlations. In the extreme case of disorder uncorrelated on the sites of lattice, it is easy to grasp that to avoid the traps is hardly possible beyond the linear sizes of several lattice constants. These preliminary considerations 
require a lot of further research which is computationally very demanding.
Nevertheless, the insights obtained from simplified 1d models are very important. They can drive and strongly impact the follow-up research, as it has  been multiply proven in the historical development of the theory of both normal and anomalous diffusion.

\section{Summary and conclusions}

In this work, we studied numerically viscoelastic subdiffusion governed by a fractional Langevin equation in stationary Gaussian random potentials for several models of decaying correlations.
Such theoretical models are of special interest in the context of biologically relevant viscoelastic subdiffusion in random environments. 
 Our study revealed several surprises. First of all, on the ensemble level the influence of random potential is almost completely negligible for $\sigma = k_BT$. Viscoelastic subdiffusion easily wins over the potential randomness, even if the potential is wildly fluctuating as in the case of exponentially decaying correlation, see in Fig. \ref{Fig1}. This is a very unexpected result because (i) normal diffusion in such potentials is suppressed by the factor $\exp[-(\sigma/k_BT)^2]$ which is approximately $0.368$ in this  case, (ii) slowness combined with sluggishness intuitively should result into a super-slowness. However, this intuition fails completely. Nevertheless, this surprising result was already partially anticipated in view of akin influence of the periodic potentials on viscoelastic subdiffusion \cite{GoychukPRE09,GoychukACP12}. It has precisely the same explanation:  Distributions of the escape times out of metastable minima have finite moments, and the asymptotic behavior is determined by viscoelastic long-time correlations in the medium that yield unobstructed subdiffusion. With the increase of the disorder strength to $\sigma =2 k_BT$, the presence of a transient behavior becomes slightly visible. However, on the ensemble level the effect is really weak and one can clearly deduce from numerics that the asymptotic regime is almost achieved at the end of our simulations. At odds with $\exp[-(\sigma/k_BT)^2]\approx 0.018$ for normal diffusion in this case, viscoelastic subdiffusion is practically not suppressed at all. Nevertheless, in spite of a barely noticeable effect on the ensemble level, single-trajectory averages exhibit a substantial scatter. This can provide a key insight to understand some experiments on subdiffusion in biological cells. With a further increase of randomness strength to $\sigma\sim (5 - 10)k_BT$, a very distinct behavior emerges. For $\sigma =10\; k_BT$, it is clearly Sinai subdiffusion, $\langle \delta x^2(t)\rangle 
\propto \ln^4 (t/t_0)$, for both exponential and power law correlations. Its origin can be explain in a very similar manner as in the case of normal diffusion in such potentials.
It is caused by extremal value fluctuations of the potential $\delta U_{\rm max}(x)\propto \sqrt{\ln x}$ and
has clearly a transient character. Ultimately, the regime of potential-free viscoelastic subdiffusion will be reached.  However, the transients can last so long that they never will be reached in reality. For an intermediate $\sigma=5\; k_BT$, a more complex behavior in Eq. (\ref{central}) substantiate, in agreement with numerics, the reasoning based on a scaling argumentation \cite{Goychuk2017} and Arrhenius character of viscoelastic diffusion over potential barriers \cite{GoychukPRE09}. It can, however, be also described with some effective power law exponent $\alpha(t)$ that temporally can be nearly a constant, which can be confused with a CTRW subdiffusion \cite{Goychuk2017}.

The author is confident that these highly surprising results will attract attention of both theorists and experimentalists leading to a further research in this currently least explored area of anomalous diffusion. \\

\section*{Conflicts of interest}
There are various conflicts of interest with competitors in the field of anomalous diffusion and its application to biological systems.

\section*{Acknowledgment} 
Funding of this research by the Deutsche Forschungsgemeinschaft (German Research Foundation), Grant GO 2052/3-1 is gratefully acknowledged.

\appendix

\section{Fractional Fokker-Planck dynamics in Gaussian disordered potentials }\label{appendA}

Let us consider a fractional dynamics governed by the fractional Fokker-Planck equation (FFPE) \cite{Metzler99,Metzler01,GoychukPRE06Rapid,HeinsaluPRE06}
\begin{eqnarray}\label{FFPE}
\frac{{\partial }^\alpha P(x,t)}{{\partial }t^\alpha} =  D_{\alpha} \frac{\partial }{\partial
x} \left ( e^{-\beta U(x)} \frac{\partial}{\partial x} \, e^{\beta
U(x)} P(x,t)  \right )\;\; 
 \end{eqnarray} 
for the probability density $P(x,t)$ of particles in a potential $U(x)$. Here, like in Eq. (\ref{GLE1}), 
$\partial^\alpha P(x,t)/\partial t^\alpha$ denotes fractional Caputo time derivative, $\beta=1/(k_BT)$ is inverse temperature, and $D_\alpha=k_BT/\eta_\alpha$ is fractional diffusion coefficient, $0<\alpha<1$.
It must be emphasized  that FFPE (\ref{FFPE}) is not a probability density counterpart of FLE (\ref{GLE1}) in the particular case $\eta_0=0$, contrary to what was stated e.g. in an important review \cite{Zaslavsky02}. Wrong assertions of this kind are the reason for co-existence of two very different fractional dynamics, both named fractional in the literature despite fundamental differences between them, as it has been clarified later on, see e.g. in Ref. \cite{GoychukFractDyn11}. The reader should be warned of this.
Fractional dynamics in the transport direction $x$ can result from an extra (infinite) orthogonal dimension, 
like in the comb model \cite{AvrahamHavlin}, where it leads to $\alpha=1/2$ in the transport direction $x$. We impose a periodic boundary condition,  $U(x+L)=U(x)$, with a very large spatial period $L$, which is a standard trick in treating disordered systems \cite{Bouchaud1990}. In the case of diffusion of regulatory proteins on a circular bacterial DNA it is even natural. By applying in addition a constant biasing force $f_0$, 
$U(x)\to V(x)=U(x)-f_0 x$ it has been shown in Ref. \cite{GoychukPRE06Rapid,HeinsaluPRE06}, that the mean displacement of particles follows $\langle \delta x(t)\rangle =\bar v(f_0) t^{\alpha}/\Gamma(1+\alpha)$, with mean subvelocity
\begin{eqnarray}\label{stratSUB}
\bar v_{\alpha}(f_0) = \frac{ D_{\alpha} L \, [1 - \exp(-\beta
f_0 L)]}{\int_{0}^L \mathrm{d} x \int_{x}^{x+L} \mathrm{d} y \,
\exp(-\beta[V(x) - V(y)])}, \, 
\end{eqnarray}
which generalizes a famous result by Stratonovich \cite{Strato58,StratoBook,RiskenBook} towards FFPE dynamics in arbitrary periodic potentials. The corresponding biased diffusion obeys a universal scaling relation 
\cite{GoychukPRE06Rapid,HeinsaluPRE06}
\begin{eqnarray} \label{anomal}
\lim_{t \to \infty} \frac{\langle \delta x^2(t) \rangle}{
\langle \delta x(t) \rangle^2} =\frac{2
\Gamma ^2(\alpha + 1)}{\Gamma(2 \alpha + 1)} - 1\;,
\end{eqnarray}
reflecting stochastic properties of underlying random clock governing this type of fractional subdiffusion  \cite{GoychukFractDyn11}. The unbiased subdiffusion, $f_0\to 0$, obeys 
$\langle \delta x^2(t)\rangle_{f_0=0} =2\bar D_\alpha t^{\alpha}/\Gamma(1+\alpha)$, with a potential-renormalized anomalous diffusion coefficient \cite{HeinsaluJP07,GoychukFractDyn11},
\begin{eqnarray} \label{FLJ}
\bar D_\alpha = \frac{ \,D_\alpha L^2}{
\int_0^L \mathrm{d}x \, \exp\left[\beta U(x)\right] \int_0^L
\mathrm{d}y \, \exp\left[-\beta U(y)\right] }, \,
 \end{eqnarray}
which follows from (\ref{stratSUB}) upon use of the fluctuation-dissipation theorem, or linear response 
relation, $\bar D_\alpha=\lim_{f_0\to 0}k_BT \bar v_\alpha(f_0)/f_0$. It generalizes a celebrated result by Lifson and Jackson \cite{LifsonJackson} towards fractional FFPE diffusion in periodic potentials. 
The results in Eqs. (\ref{stratSUB})-(\ref{FLJ}) were verified in extensive numerical simulations for various periodic potentials \cite{GoychukPRE06Rapid,HeinsaluPRE06,HeinsaluJP07} and are firmly established.
Using a spatial average, $\overline{ C_L^{\pm}}=(1/L)\int_0^{L}w_{\pm}(x)dx$, of statistical weight-functions $w_{\pm}=\exp[\pm \beta U(x)]$, the result in (\ref{FLJ}) can be rewritten as 
$\bar D_\alpha = D_{\alpha}/[\overline{ C_L^{+}}\;\overline{ C_L^{-}}]$. Furthermore, assuming ergodicity of the random process 
$w_{\pm}(x)$ in space, one can replace in the limit $L\to\infty$ the spatial averages with the corresponding  ensemble averages and for Gaussian disorder we obtain the central result of this Appendix:
\begin{eqnarray}
\langle \delta x^2(t)\rangle =2D_\alpha e^{-\sigma^2/(k_BT)^2}t^{\alpha}/\Gamma(1+\alpha).
 \end{eqnarray}
Following \cite{GoychukPRL14} it is easy to show that this result is valid asymptotically for any stationary Gaussian potential with decaying correlations.
This result generalizes the normal-diffusion result by Zwanzig \cite{ZwanzigPNAS} \textit{et al.} towards FFPE subdiffusion in such potentials. However, it is not applicable to viscoelastic fractional  
subdiffusion studied in this paper. 

\section{Velocity autocorrelation function for nanoparticles in viscoeleastic media and  neglection of inertial effects }\label{appendB}

In this Appendix, a justification of the neglection of inertial effects is given. For free 
viscoelastic subdiffusion with the inertial effects included, the normalized stationary velocity autocorrelation function (VACF) in the Laplace space reads, see Eq. (B26) in Ref. \cite{GoychukACP12} 
or Eq. (A3) in Ref. \cite{GoychukPRE09},
\begin{eqnarray}\label{Kv}
\tilde K_v(s)=\frac{1}{s+\tilde \eta(s)/m},
\end{eqnarray}
where $\tilde \eta(s)$ is Laplace-transform of a general memory kernel and $m$ is the mass of particles. For the memory kernel in this paper it is
\begin{eqnarray}\label{Kv2} 
\tilde K_v(s)=\frac{1}{s+\gamma_0+\gamma_\alpha s^{\alpha-1}}\;,
\end{eqnarray}
where $\gamma_0=\eta_0/m$ and $\gamma_\alpha=\eta_\alpha/m$. In this case, $\int_0^\infty K_v(t)dt=0$, 
for any $\eta_\alpha\neq 0$.
For normal diffusion with $\eta_\alpha=0$, $K_v(t)=\exp(-t/\tau_v^{(0)})$, in the time domain, where $\tau_v^{(0)}=m/\eta_0$ the velocity relaxation constant. For the case of purely fractional friction with $\eta_0=0$, $K_v(t)=E_{2-\alpha}[-(t/\tau_v)^{2-\alpha}]$ \cite{GoychukACP12,LutzFLE}, where $E_a(z)=E_ {a,b=1}(z)$ is the Mittag-Leffler function, and $\tau_v=(m/\eta_\alpha)^{1/(2-\alpha)}$ is the velocity relaxation constant of anomalous relaxation, which has  in this case a negative power-law tail, $K_v(t)\propto -1/(t/\tau_v)^{2-\alpha}$, for $t\gg \tau_v$. Notice, that in both cases the velocity relaxation is instant in the strict limit $m\to 0$. This limit is singular because the mean-square thermal velocity diverges 
$v_T^2=k_BT/m\to \infty$. This the reason why both the normal Brownian motion (Wiener process) and fBm are not differentiable, being singular processes in this respect, though of a wide and common use in physics. Furthermore, after some lengthy algebra one obtains the following exact result in the case $\alpha=1/2$:
\begin{eqnarray}\label{Kvexact}
K_v(t)=\sum_{i=1}^3 c_i z_i^2\exp(z_i^2 t/\tau_v){\rm erfc}(z_i\sqrt{t/\tau_v})
\end{eqnarray}
where $\tau_v=(m/\eta_\alpha)^{2/3}$,
\begin{eqnarray}
z_1 &= &\frac{2}{3\cdot 4^{1/3}}\frac{\gamma_0\tau_v}{(\sqrt{1+12(\gamma_0\tau_v)^3/81}-1)^{1/3}}\nonumber\\
&-& \frac{4^{1/3}}{2}(\sqrt{1+12(\gamma_0\tau_v)^3/81}-1)^{1/3}, \\
z_{2,3}& = &-z_1/2\pm i\frac{\sqrt{3}}{2} \nonumber \\
&\times& \Big [\frac{4^{1/3}}{2}(\sqrt{1+12(\gamma_0\tau_v)^3/81}-1 )^{1/3}\nonumber \\
&+&\frac{2}{3\cdot 4^{1/3}}\frac{\gamma_0\tau_v}{(\sqrt{1+12(\gamma_0\tau_v)^3/81}-1 )^{1/3}}\Big ]
\end{eqnarray}
and 
\begin{eqnarray}
c_1 &= &\frac{1}{(z_1-z_2)(z_1-z_3)},\nonumber \\
c_2 &= &\frac{1}{(z_1-z_2)(z_3-z_2)},\nonumber \\
c_3 &= &\frac{1}{(z_2-z_3)(z_1-z_3)}\;.
\end{eqnarray}
For $\gamma_0=0$, $z_1=1$ and $z_{2,3}=-1/2\pm i\sqrt{3}/2$.
Independently of $\gamma_0$, the asymptotics reads
\begin{eqnarray}
K_v(t)\sim - \frac{1}{2\sqrt{\pi}(t/\tau_v)^{3/2}}
\end{eqnarray}
for $t\gg \tau_v$. Notice that already for $t>100\tau_v$, $K_v(t)$ is really small and the overdamped approximation can be accurately used, on theoretical grounds. In fact, however, the initial ballistic regime is already
over for $t>(3 - 5)\tau_v$ in reality, see e. g. Figs. 5, 6 in Ref. \cite{GoychukACP12}. It is important to emphasize that this fact does not contradict to another fact that double integration of $v_T^2K_v(t)$ yields the mean square displacement whose behavior is determined by the discussed tail and where the time scale $\tau_v$ drops out. This important time scale defines the initial ballistic regime, see in Refs. \cite{GoychukPRE09, GoychukACP12} which is of no importance in this work.
For $0\leq \gamma_0\tau_v<  \gamma_0^{(c)}\tau_v \approx 0.220404$, $K_v(t)$ changes its sign precisely three times. For a larger $\gamma_0\tau_v$, a single change of the sign occurs.

Let us estimate now the time scale $\tau_v$  of the velocity relaxation 
for diffusion of nanoparticles in viscoelastic cytosol of living cells.
For the colloidal gold particles with radius $R=2.5$ nm in Ref. \cite{Guigas} subdiffusion was found in HeLa cells with $\alpha\approx 0.51\approx 0.5$ and $D_\alpha\approx  0.644 \;{\rm \mu m^2/s^\alpha} $ (see in Table I in Ref. \cite{PRE12b}) on the time scale up to 1 s, at least. Let us use this experimental value and the generalized Einstein relation $D_\alpha=k_BT/\eta_\alpha$ to estimate $\tau_v=\left (4\pi \rho R^3D_{1/2}/(3k_BT)\right)^{2/3}$, where $\rho$ is the mass density of particle. With the gold mass density $\rho =19.3$ kg/m$^3$ and  room $k_BT=4.1{\rm \;pN\cdot nm}$, we obtain $\tau_v\approx 28.08$ ps (picoseconds!). For lighter particles such as globular proteins with $\rho \sim 1.2$ kg/m$^3$ it will be even smaller.  Thus,  already on the time scale of nanoseconds and larger the inertial effects are completely negligible, for sure. Furthermore, the normal diffusion coefficient $D_0$ in the water component of cytosol can be estimated using the Einstein-Stokes relation as $D_0=k_BT/(6\pi\zeta_w R)$, where $\zeta_w \approx 1\; {\rm mPa\cdot s}$ is water viscosity. This yields $D_0\approx 87 \;{\rm \mu m^2/s}$. The corresponding time $\tau_0$ separating the initially normal diffusion and subdiffusion in our paper is hence $\tau_0=(D_{1/2}/D_0)^2\approx 54.77\;\mu$s. It is of the order $10^6$ larger than $\tau_v$, which once again confirms that the inertial effects are completely negligible. In fact, the time step in our numerical simulations exceeds this estimated $\tau_v$ by orders of magnitude. Furthermore, the scaling time $\tau_r$ used in simulations can be expressed for $\alpha=1/2$ via dimensionless $\tilde \eta_0$ as $\tau_r=\tau_0/\tilde \eta_0^2$, which for $\tilde \eta_0=0.1$ used in numerics gives $\tau_r=100\tau_0\approx 5.48 $ ms. Hence, the maximal time $t_{\rm max}=2\times 10^5$ in our numerics corresponds to about $t_{\rm max}=1096$ s. The corresponding length scale, which we took to be the correlation scale, is estimated as $\lambda=\tau_r^{1/4}(D_{1/2}\sigma/k_BT)^{1/2}\approx 0.218 \sqrt{\sigma/k_BT}$  $\mu$m, i.e. $\lambda \approx  0.308$ $\mu$m for $\sigma=2$ (in units of $k_BT$), or $\lambda \approx 0.689 $ $\mu$m for $\sigma=10$, which are typical mesh sizes in random actin meshworks \cite{WongPRL}.

\subsection{Overdamped approximation in a potential}

Also for dynamics in random potentials one can show that the overdamped approximation works typically 
well for colloidal particles in viscoelastic fluids.
Random potentials considered in this paper are locally piece-wise parabolic. Therefore, it compels to consider relaxation in a parabolic potential with curvature $\kappa$.
The corresponding position relaxation function, which coincides with  normalized stationary autocorrelation function of the coordinate \cite{GoychukACP12}, for the model considered reads, 
\begin{eqnarray}\label{theta} 
\tilde \theta(s)=\frac{s+\gamma_0+\gamma_\alpha s^{\alpha-1}}{s^2+\gamma_0 s+\gamma_\alpha s^{\alpha}+\omega_0^2}\;,
\end{eqnarray}
in the Laplace-space,
cf. Eq. (B7) in  Ref. \cite{GoychukACP12}. Here, $\omega_0=\sqrt{\kappa/m}$ is vibrational frequency. To clarify whether relaxation dynamics is overdamped or not it is convenient to scale time in the units of $1/\omega_0$, and $s\to \tilde s=s/\omega_0$. Then, the correspondingly scaled, relevant non-dimensional frictional constants read, $\bar \gamma_0=\gamma_0/\omega_0$, and $\bar \gamma_\alpha=1/(\omega_0\tau_v)^{2-\alpha}$, where $\tau_v$ is the above time constant of a potential-free anomalous relaxation of velocity. It is absolutely clear that the dynamics is overdamped for $\bar \gamma_0>2$ when $\bar \gamma_\alpha=0$. It is also beyond doubts that it will be overdamped for a smaller value of $\bar \gamma_0$ in the case when an additional memory friction is present, especially 
if it is overdamped already for $\gamma_0=0$ due to the memory friction alone.
 Hence, let us first estimate 
$\gamma_0$. For a gold sphere of radius $2.5$ nm in water we have: $\eta_0\approx 4.713\cdot 10^{-11}$ kg/s, and 
$m\approx 1.263 \cdot 10^{-24}$ kg. Hence, $\gamma_0\approx 3.731 \cdot 10^{13}$ 1/s. Therefore, for any frequency $\omega_0$ lower that $10^{13}$ 1/s, or about $300$ cm$^{-1}$ in spectroscopic units, which is a typical frequency of molecular vibrations, the dynamics of such a colloidal particle in a trapping potential is clearly overdamped.
Let us estimate this frequency in a worse case scenario of a singular disorder corresponding to exponentially decaying correlations in space. 
The disorder is regularized by considering it on a lattice with the grid constant $\Delta x$  \cite{GoychukPRL14,Goychuk2017}. We can estimate a typical value of 
$\kappa$ from the condition $\langle \delta f^2(0)\rangle^{1/2}\sim \kappa \Delta x$, where 
$\langle \delta f^2(0)\rangle^{1/2}=\sqrt{2}\sigma/\sqrt{\Delta x\lambda }$
 is the rms amplitude of the random force fluctuations \cite{Goychuk2017}. This yields an estimate
 $\kappa \sim 2^{1/2}\sigma/[(\Delta x)^{3/2}\lambda^{1/2}]$.
With $\lambda=308$ nm and $\Delta x=0.02\lambda=6.16$ nm, for $\sigma=2\;k_BT_{r}=8.2\;{\rm pN \cdot nm}$, we have $\kappa\approx 0.0432$  pN/nm, which for $m=1.263\cdot 10^{-24}$ kg yields $\omega_0=\sqrt{\kappa/m} \sim  5.85 \cdot 10^{9}$ 1/s. The corresponding $\bar \gamma_0\approx 6377.8$ is really large. Let us estimate also $\bar \gamma_\alpha$ in this case.
With $\tau_v=28.08$ ps (see above) we obtain $\bar \gamma_\alpha\approx 15.02$. For such a large $\bar \gamma_\alpha$,
dynamics is clearly overdamped even for $\bar \gamma_0=0$. Indeed,
the case with $\gamma_0=0$ for arbitrary $\alpha$ was studied in Refs. \cite{BurovPRL08,BurovPRE08}. It has been shown therein that for $\alpha=1/2$, the relaxation is overdamped and monotonous at $\omega_0<\omega_m=0.426$ in the units where $\tau_v=1$. This critical condition translates into
 $\bar \gamma_\alpha > \gamma_m=3.596$, in our notations. 
Clearly, inertial effects are completely negligible from all points of view. 
Indeed, motion of colloidal particles can be treated typically as overdamped already due to the water component
of such complex viscoelastic media as polymeric solutions and cytosol.
In the case of power law potential correlations, subdiffusion will be overdamped for a much wider range of parameters.
Indeed, in this case a typical $\kappa\sim 2\sigma/\lambda^2$ is much smaller than
in the case of singular disorder.
This indicates that overdamped approximation can be justified in many cases and it presents an important model of general interest. 

\section{Numerical algorithm }\label{appendC}

Numerical algorithm to propagate the stochastic dynamics in Eq. (\ref{embedding1}), (\ref{embedding2}) is based on the well-known stochastic Heun algorithm, or the second order stochastic Runge-Kutta method \cite{GardBook}. 
Below we sketch it. First, we rewrite (\ref{embedding1}), (\ref{embedding2}) as 
\begin{eqnarray}
  &&\dot{x}=F(x,x_i)+\sqrt{2D_0}\zeta_0(t), \nonumber \\
  &&\dot{x}_i=\nu_i(x-x_i)+\sqrt{2D_i}\zeta_i(t),
\end{eqnarray}
with $D_0=k_BT/\eta_0$, $D_i=k_BT/\eta_i$, and $F(x,x_i)=[f(x)-\sum_{i=1}^N k_i(x-x_i)]/\eta_0$.
On each integration time step $\Delta t$ one generates anew $N+1$  independent zero-mean Gaussian variables $W_i$ with unit variance, $i=0,1,2...N$ (Mersenne Twister pseudo-random number generator was used for this). Each propagation step in the discretized time dynamics, $x_k=x(k\Delta t)$, $x_{i,k}=x_i(k\Delta t)$, from $t_k=k\Delta t$ to $t_{k+1}=t_k+\Delta t$ consists of two substeps. In the first substep,
\begin{eqnarray}
  &&x_k^{(1)}=x_{k}+F(x_k,x_{i,k})\Delta t+\sqrt{2D_0\Delta t} W_0, \nonumber \\
  &&x_{i,k}^{(1)} =x_{i,k}+\nu_i(x_k-x_{i,k})\Delta t+\sqrt{2D_i \Delta t} W_i\;.
\end{eqnarray}
In the second (final) step,
\begin{eqnarray}
  x_{k+1}&=&x_{k}+\left [F(x_k,x_{i,k})+F(x_k^{(1)},x_{i,k}^{(1)})\right ]\Delta t/2\nonumber \\ &+&\sqrt{2D_0\Delta t} W_0,  \\
  x_{i,k+1} &=&x_{i,k}+\nu_i\left [x_k+x_k^{(1)}-x_{i,k}-x_{i,k}^{(1)})\right ]\Delta t/2 \nonumber \\
  & + &\sqrt{2D_i \Delta t} W_i\;. \nonumber
\end{eqnarray}
Notice that $W_i$ must be the same numbers on the both substeps \cite{GardBook}. The algorithm was implemented in CUDA and propagated in parallel (many different particles with different initially random preparations at the same time) on GPU processors.

\bibliographystyle{PRX}

\bibliography{aps1}

\end{document}